# Conflict and Congruency Effects in Large Language Models: In-Weight and In-Context Competition in a Verbal Conflict Task

Xiaoyang Hu[1], Mike Angstadt[2], Shane Storks[3], Zan Huang[2], Aman Taxali[2], Alex Weigard[2], Richard L. Lewis[3,4,5], Chandra Sripada[2,3,6,*]

1. Department of Computer Science, Brown University
2. Department of Psychiatry, University of Michigan
3. Weinberg Institute for Cognitive Science, University of Michigan
4. Department of Psychology, University of Michigan
5. Department of Linguistics, University of Michigan
6. Department of Philosophy, University of Michigan

* To whom correspondence should be directed: sripada@umich.edu

## Abstract

Congruency effects, observed in conflict tasks such as Stroop and flanker tasks, have been investigated for nearly a century in psychology and neuroscience, but their mechanistic basis is not fully understood. We introduce a verbal-only LLM conflict task in which a prompt stem elicits a default same-color completion and an explicit rule either agrees with (congruent condition) or conflicts with (incongruent condition) the completion. Gemma-2-2B and six Pythia models ranging from 410M to 12B parameters showed strong default same-color tendencies, and six of seven models showed strong congruency effects. Using causal attribution analysis, attention analysis, and attention ablations, we identified distinct processing pathways in these LLMs: a pathway involving short-range attention to a superficial color cue that is preferentially activated in the congruent condition, and a pathway involving long-range attention to the rule prefix that is preferentially activated in the incongruent condition. Fine-tuning that strengthened the default same-color tendency had divergent effects on task conditions, reducing incongruent performance while increasing congruent performance. In contrast, increasing rule set size selectively impaired incongruent performance. These converging findings support an account in which congruency effects in this task arise from competition between an in-weight default mapping and an in-context rule-based mapping. More broadly, our findings illustrate how LLMs can serve as model systems for mechanistic analysis of competition between default and rule-governed response tendencies within a single learned network.

## Introduction

The ability to override an automatic or prepotent response in favor of a context-appropriate one is central to adaptive behavior (Miller and Cohen 2001), and its failure is implicated in impulsivity (Bari and Robbins 2013), addiction (Goldstein and Volkow 2011), and attentional disorders (Barkley 1997). A canonical manifestation is the congruency effect, in which conflict between task-irrelevant and task-relevant information incurs a performance cost. Observed across many “conflict task” paradigms, most famously the Stroop task, the congruency effect is one of the most widely studied and replicated phenomena in experimental psychology (Stroop 1935; Simon 1969; Eriksen and Eriksen 1974; MacLeod 1991).

While much has been written about congruency effects in conflict tasks, the mechanistic basis of processing during these tasks is not yet fully understood. According to “dual-process theories”, conflict tasks implicate two qualitatively distinct systems or processes: a fast, automatic system that produces prepotent responses, and a slower, deliberate system that engages to override these responses based on contextual demands (Posner et al. 2004; Shiffrin and Schneider 1977; Sloman 1996; Evans 2008; Kahneman 2011). A frequent criticism of these accounts, however, is that they are underspecified: a “process” or “system” too often involves labeling behavioral dissociations rather than specifying step-by-step algorithms or mechanisms.

The parallel distributed processing (PDP) account is one important approach that attempts to fill in more mechanistic details. In an influential work, Cohen, Dunbar, and McClelland (1990) showed that Stroop performance can be captured by a network with two pathways (color naming and word reading) of differing strengths. Their model additionally includes attention control nodes that modulate the strength of either pathway. The automatic/controlled distinction emerges in their model as graded competition between these pathways under attentional modulation.

While this PDP model, and related extensions of the model from other theorists (Botvinick et al. 2001; Herd et al. 2006; Kalanthroff et al. 2018), are mechanistically more precise, this family of models too faces several limitations. First, these models incorporate attention control as a primitive within the model. For example, in the Cohen, Dunbar, and McClelland (1990) model, the effect of following novel color naming task instructions is captured within the model in terms of activation at an attention control node. The model itself provides no further details on how verbal task instructions are interpreted, task representations are constructed, and how they are brought to bear as signals that compete with automatic response tendencies. Second, the separate pathways for color naming, word reading, and top-down attention control are built into the model by the researchers. The model is thus not able to answer questions about whether separate response pathways and top-down control would emerge in a data-driven way from general-purpose learning, or whether they necessitate specific architectural pathways that must be pre-installed.

In recent years, large language models (LLMs) have emerged as an intriguing testbed for computational hypotheses about cognitive phenomena. Recent studies demonstrate robust congruency effects across a number of conflict tasks and a wide range of language models (Luo et al. 2025), as well as related phenomena such as trial-to-trial conflict adaptation (X. Hu 2025). These findings are remarkable because LLMs were trained not on narrow tasks but rather by general-purpose methods, specifically predict-the-next-token self-supervised learning, on large corpora of internet text. The transformer architecture thus does not hard-code separate automatic responses and control structures, so if this structure is present in the model, it must arise from prediction pre-training.

While the presence of congruency effects in LLMs across a range of conflict tasks is now well established, the internal processing patterns within LLMs that produce these effects remain poorly understood. One way to address this gap is through application of the tools of mechanistic interpretability (Olah et al. 2020; Elhage et al. 2021; Olsson et al. 2022; Wang et al. 2022; Lindsey et al. 2025; Ameisen et al. 2025), which allow researchers to assess and quantify the circuits and processing patterns that contribute to model responses. Specific mechanistic interpretability methods enable researchers to (among other things): build sparse encoders and transcoders to identify specific features implicated in model responses; construct causal attribution graphs that estimate causal effects of individual features on other features and on model responses; inspect and ablate attention heads to assess what the model is attending to during key processing steps and what causal effects these heads have on processing; and fine-tune models in order to enhance (or mitigate) specific dimensions of model performance. As far as we know, the application of mechanistic interpretability techniques such as these to conflict tasks in LLMs to examine the underlying mechanisms and processing patterns that produce observed congruency effects has not yet been undertaken.

In addition to yielding specific mechanistic insights, LLMs may offer new theoretical resources for understanding the "dual processes" long thought to underlie conflict tasks and congruency effects. It is now widely recognized that LLMs can acquire novel task mappings specified in context itself, either through explicit natural language instructions or through presentation of a few examples (Radford et al. 2019; T. Brown et al. 2020). This motivates a now well-recognized distinction between two sources of control over an LLM's response: 1) an "in-weight" contribution, which relies on a mapping already stored in the model's parameters (learned previously through gradient descent) and which is merely cued by the prompt; and 2) an "in-context" contribution, which relies on a mapping explicitly given in the prompt, which must be interpreted and applied online.

For example, in "Translate from French to English: chat ->", a tendency to answer cat reflects an in-weight contribution, since the relevant French-English mapping is already stored in the model's parameters and merely cued by the prompt. By contrast, if the prompt first states "In this toy language, zarp means cat", and then the model is next asked to "Translate zarp to English", then answering cat here reflects an in-context contribution, since the relevant mapping is prompt-specified and applied online. Importantly, the distinction is not between two different physical mechanisms, since both are implemented within the transformer neural network in the same forward pass, but between two different sources of task-relevant mappings: one latent in the weights and cued by the current input, the other explicitly specified in the prompt and implemented through ongoing computation.

The in-weight/in-context distinction which arose with LLMs has been fruitfully used to illuminate key phenomena in human cognitive science, including curriculum effects in learning (Russin et al. 2025; Pesnot Lerousseau and Summerfield 2026), temporal contiguity effects in episodic memory recall (Ji-An et al. 2024), and asymmetric belief updating in instrumental tasks (Schubert et al. 2024). Another cognitive science direction in which this distinction might be fruitfully applied is understanding the etiology of congruency effects in conflict tasks. Theorists have long characterized these tasks as implicating an "automatic" or "prepotent" tendency (which can naturally be understood in terms of an in-weight tendency) which competes with the novel instructions provided in the task (which are naturally understood as implicating in-context processing) (MacLeod 1991; Miller and Cohen 2001). As far as we know, however, the hypothesis that congruency effects arise from analogs of in-weight/in-context competition in human cognition has not previously been systematically investigated.

In the present study, we examine the processing patterns in LLMs during a conflict task using a number of mechanistic interpretability tools. While previous studies have directly presented standard conflict tasks, such as the Stroop and flanker tasks, to multimodal vision language models, we instead study a novel "verbal-only" conflict task. We adopt this approach because in multimodal vision language models, many mechanistic interpretability methods are either not available (e.g., causal attribution graphs), underdeveloped (sparse autoencoders and transcoders for visual features), or challenging to interpret (fine-tuning methods that selectively or differentially impact modalities).

We thus introduce the verbal-only "crayon conflict task", which consists of a prompt *stem*, which generates an automatic tendency for a color word completion and a *rule-based prefix*, which either agrees with (congruent condition) or competes with (incongruent condition) the completion supported by the stem. We demonstrate that absent the rule-based prefix (i.e., a stem-only prompt), all seven models studied, spanning 410M to 12B parameters, show the expected strong default response tendencies. When the rule-based prefix is appended, we find six of seven models show robust congruency effects in which probabilities for correct responses (relative to incorrect responses) are significantly higher in congruent trials compared to incongruent trials. We apply a number of mechanistic interpretability methods to illuminate processing patterns in this task, including causal attribution analysis, attention analysis, attention ablations, fine-tuning manipulations, and rule set size manipulations. These methods identify distinct processing pathways that are preferentially activated in the congruent and incongruent conditions, and they provide initial intriguing evidence linking congruency effects to competition between in-weight and in-context processing.

# Methods

## 1. Task

We study a novel verbal-only conflict task that we dub the “crayon” task. The general form of the prompts and key terms are shown in Figure 1.

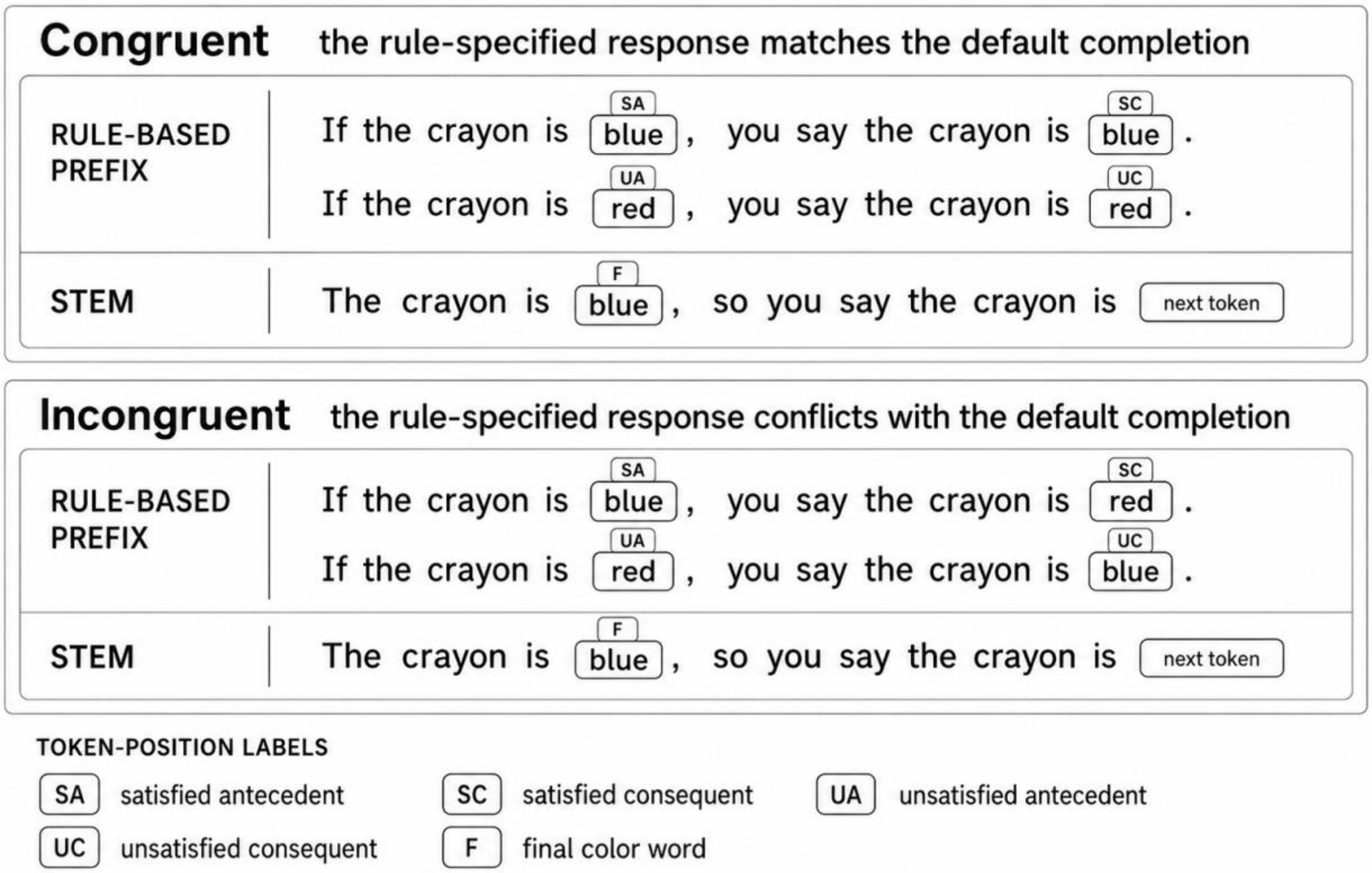

**Figure 1**: *Example prompts and key terms*.

We refer to the first and second sentences of the prompt as the “prefix” and the last sentence of the prompt as the “stem”. In the congruent condition, the rule in the prefix is familiar while in the incongruent condition, the rule is an unfamiliar mapping in which if the crayon has a certain color, you say it has a different color.

There are five color words in each prompt. We refer to the color word in the stem as the final color word (F). The prefix contains two conditional clauses, i.e., if-then rules. The clause that contains the antecedent that is satisfied by the final color word contains the satisfied antecedent (SA) and satisfied consequent (SC), while the other clause contains the unsatisfied antecedent (UA) and unsatisfied consequent (UC).

We used the color word set {blue, red, green, brown, yellow}, which we verified are single tokens in all seven models studied. These five colors generated 20 ordered color pairs. Crossing these with two congruency conditions and two satisfied-clause positions yielded 80 prompts, 40 congruent and 40 incongruent. We present each prompt to each model independently, without a system prompt or conversation history.

## 2. Models

Our main analyses are conducted on Gemma-2-2B (Gemma Team et al. 2024). We additionally study a series of six Pythia models (Biderman et al. 2023) ranging from 410 million parameters to 12 billion parameters to assess robustness of results and sensitivity to systematic variation in model size. Models were downloaded from Hugging Face.

## 3. Metrics

We examine how task conditions, including congruency versus incongruency, fine-tuning, and rule set size, affect model performance. Following previous work (Wang et al. 2022; Binz and Schulz 2023; Lampinen et al. 2024; Jones et al. 2024; J. Hu et al. 2025), we quantify performance using *Δ probability* and *Δ log probability*, measures of the model's relative strength of evidence for the correct versus incorrect response. More specifically, for each prompt, let $p_c$ denote the model's next-token probability for the correct color word and $p_i$ its next-token probability for the competing incorrect color word. We compute two signed measures:

$$delta\ probability\ = \ \Delta P\ = \ p_c\ -\ p_i$$
$$delta\ log\ probability\ = \Delta log\ P\ = \ log(\,p_c)\ - log(p_i)\ = \ log(\frac{p_c}{p_i})$$

Positive values indicate that the model favors the correct response; values near zero indicate that the two responses are approximately balanced; and negative values indicate that the model favors the incorrect response.

The motivation for these metrics comes from sequential sampling, or evidence accumulation, models, the leading framework in cognitive science for modeling forced-choice tasks. In these models, relative strength of evidence favoring one response over another, often formalized in likelihood-ratio terms (Ratcliff and Rouder 1998; S. D. Brown and Heathcote 2008), is a key underlying variable that contributes to observed patterns of reaction times and accuracies (i.e., the more strongly evidence favors the correct response, the faster and more accurate responses are predicted to be). The language model's assignment of probabilities to next tokens can be interpreted as a static readout of response evidence or decision strength for each response at the point of generation, and previous work confirms that $\Delta$ log probability metrics from language models during tasks are predictive of human accuracy (Jones et al. 2024) and reaction times (Lampinen et al. 2024) during those same tasks, consistent with the assumptions of the sequential sampling framework.

We report both $\Delta$ probability and $\Delta$ log probability for all analyses. For most analyses, we focus on $\Delta$ probability in the main manuscript because it is directly interpretable on the probability scale, and we report $\Delta$ log probability in the Supplement. For all analyses, the directionality of results was identical across these metrics, and the degree of statistical significance did not meaningfully change.

## 4. Causal Attribution Graphs

We created causal attribution graphs (Lindsey et al. 2025) for each of the 80 crayon task prompts using the circuit-tracer framework for Gemma-2-2B (Hanna et al. 2025). These graphs provide a prompt-specific graphical map of processing leading to a target response. Nodes represent prompt tokens, activated features, and output tokens, and weighted directed edges estimate how strongly each node promotes or suppresses downstream features and, ultimately, the target output.

To assess differences in attribution patterns across task conditions, we calculated the total graph-based influence routed through each of the five color-word positions {SA, SC, UA, UC, F} on the correct-response logit. We report this measure separately for congruent and incongruent conditions. At the time of analysis, compatible pretrained transcoders were available for Gemma-2-2B but not for the Pythia models; accordingly, this analysis was restricted to Gemma-2-2B. GemmaScope per-layer transcoders (Lieberum et al. 2024) were used with the "gemma" shortcut in the circuit-tracer library. This uses the transcoder at each layer with the lowest mean number of active features. Each of the 80 attribution graphs was generated with a maximum of 4096 features, and the top 10 output logits. Graphs were pruned with default settings of 0.8 node influence threshold and 0.98 edge influence threshold. Links from MLP reconstruction error nodes were not included in the calculation of total graph-based influence. For each color word position, we summed the raw weights of each link starting at a node in that token position and ending at the correct output logit node.

## 5. Attention Heat Maps and Attention Ablation

We assessed differences in attention patterns from the final token to the token positions for the five color words {SA, SC, UA, UC, F}. Attention weights were averaged across heads, and we report results layerwise with a heatmap for each color word position as well as a single value summed across layers for each color word position.

We additionally performed an all-downstream-position ablation of attention to the satisfied-consequent (SC) token position. At every layer and in every attention head, we zeroed the attention connection from each token position following SC to the SC position. Thus, no subsequent token position could directly attend to SC at any layer. This intervention tests whether successful responding depends on the SC position serving as a source of information for downstream token representations. We examined the effects of the ablation on $\Delta$ probability and $\Delta$ log probability metrics separately for the congruent and incongruent conditions.

## 6. Fine-Tuning

We aimed to selectively manipulate the strength of the in-weight tendency of a model to engage in “same color” responses. This refers to the tendency of the model, when presented with the stem portion of the prompt (e.g., “The crayon is blue so you say the crayon is…”), to respond with the same color (e.g., “blue”).

Using a list of common nouns and adjectives, we generated prompts of the following three types to use as fine-tuning training data, with the target completion underlined:
I say {noun}, so you say {same noun}
The word is {noun}, so you say the word is { same noun }
The {noun} is {adjective}, so you say the {noun} is { same adjective }

We used LoRA (E. J. Hu et al. 2021) as implemented in the Parameter Efficient Fine-Tuning library (PEFT) (Mangrulkar et al. 2022). We restricted the fine-tuning to only the MLP layers of the models. Models were trained with a batch size of 16 and 4 gradient accumulation steps, so each training step represents 64 training examples. We trained for 20 steps, saving the state at each step in order to track changes in $\Delta$ probability and $\Delta$ log probability over increasing in-weight tendency training.

We examined effects of this manipulation on $\Delta$ probability and $\Delta$ log probability separately for congruent and incongruent conditions.

## 7. Rule Set Size Manipulation

We aimed to manipulate the strength of the in-context ability of the model to use the rule-based prefix to influence its responses. Prior work found that LLM rule-following deteriorates as additional rules are added and the deterioration is more pronounced for information in the middle of a longer context, i.e., a “lost in the middle effect” (Liu et al. 2024). Thus, we systematically varied the number of clauses in the prefixes from two to five clauses. Examples of congruent and incongruent prefixes with three clauses are shown below:

**Congruent**:
If the crayon is blue, you say the crayon is blue. If the crayon is red, you say the crayon is red. If the crayon is green, you say the crayon is green. The crayon is blue, so you say the crayon is

**Incongruent**:
If the crayon is blue, you say the crayon is red. If the crayon is red, you say the crayon is green. If the crayon is green, you say the crayon is blue. The crayon is blue, so you say the crayon is

We examined effects of this manipulation on $\Delta$ probability and $\Delta$ log probability separately for congruent and incongruent conditions. To assess the presence of a “lost in the middle” effect, we separated results by position of the satisfied clause (i.e., the clause with the SA and SC), that is, whether the satisfied clause occurs in the 1st, 2nd, 3rd, 4th, or 5th clause in the prefix.

# Results

Our main analysis was conducted with Gemma-2-2B. We repeated analyses with six Pythia models, and report those results in section 7 below.

**1. The crayon prompt stems (without the rule prefix) generate a strong “default” tendency to say the same color word.**

The crayon task prompt stem has the form: “The crayon is [color], so you say the crayon is”. There are five prompt stems, in each of which [color] is replaced by a member of the set {red, blue, yellow, green, brown}. We started by confirming that these prompt stems have a strong "default" tendency to yield a same-color response.

We found that for all five prompts, the next token probability for the same-color response was always the highest, with a mean probability of 0.5074 (range: 0.4219-0.5664). The mean probability for the second most likely response across the five prompts was 0.0822 (range: 0.0674-0.1064). This yielded an average *Δ* probability favoring the same-color response of 0.4252 and *Δ* log probability of 1.8211.

**2. The crayon task produces strong congruency effects**

As shown in Figure 2, we observed strong congruency effects in the crayon task. In the congruent condition, mean *Δ* probability was 0.9644, while in the incongruent condition, mean *Δ* probability was 0.5676. The difference in *Δ* probability across the two conditions was 0.3968, 95% CI [0.3529, 0.4407], $t(39) = 18.2794$, $p < 1e-8$. Accuracy was 100% in both conditions.

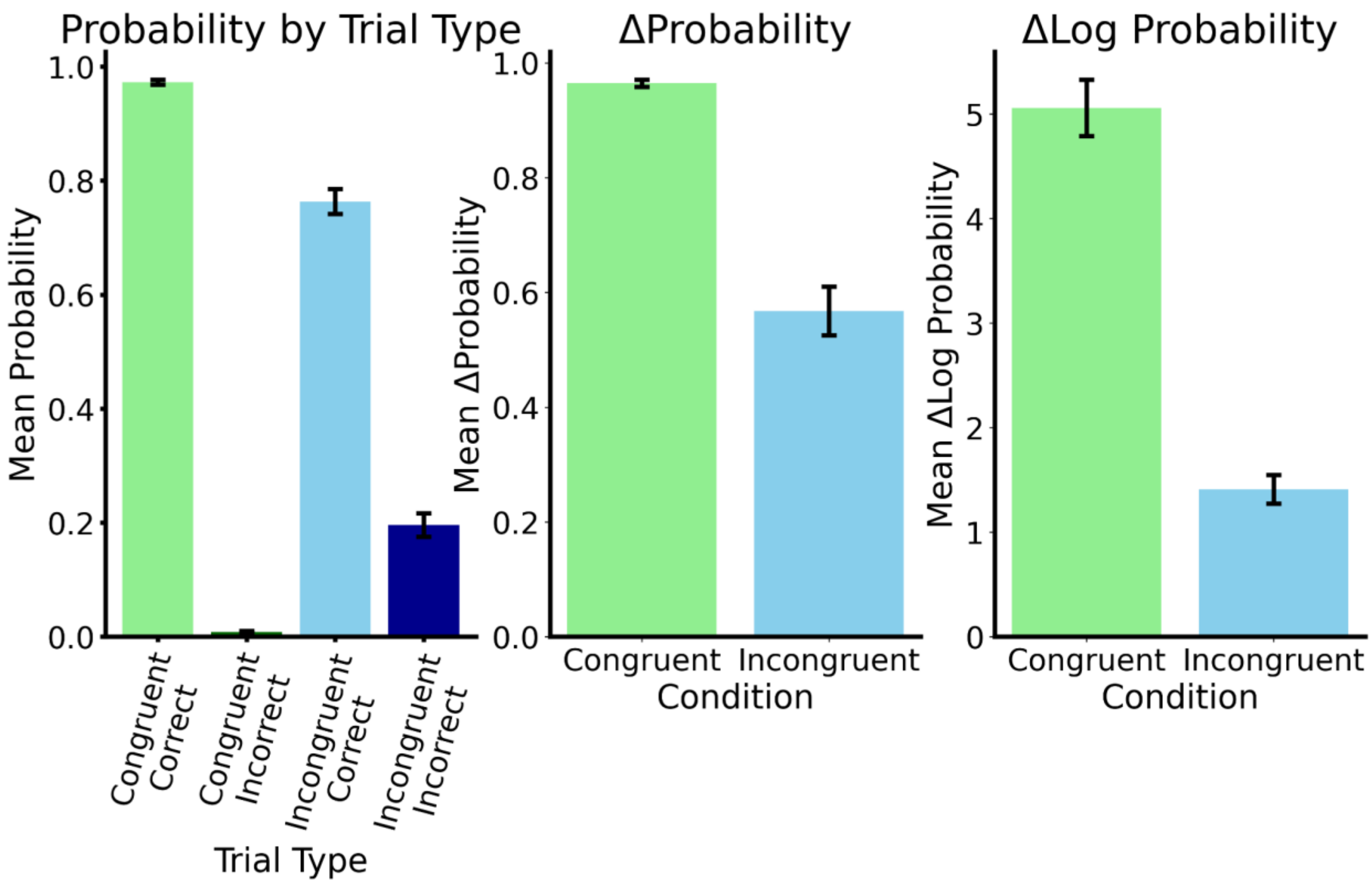

**Figure 2**: *Congruency Effects by Trial Type and Task Condition*. Strong congruency effects were observed in the crayon task, as reflected in: (Left Panel) differences in next-token probabilities for correct vs. incorrect responses in the congruent versus incongruent conditions. (Middle Panel) differences in *Δ* probability and (Right Panel) differences in *Δ* log probability. Error bars are 95% confidence intervals.

**3. There are significant differences in causal attribution outflow in congruent and incongruent conditions**

Using the causal attribution graph framework, we compared the causal geometry of information flow in the transformer in the congruent and incongruent conditions. We provide a sample visualization of causal attribution graphs for one congruent and one incongruent prompt in the Supplement, Figure S1. As shown in Figure 3, in the congruent condition, the strongest causal attribution outflow was observed at the final color word position (F). In the incongruent condition, the strongest causal attribution outflow was observed at the satisfied consequent position (SC).

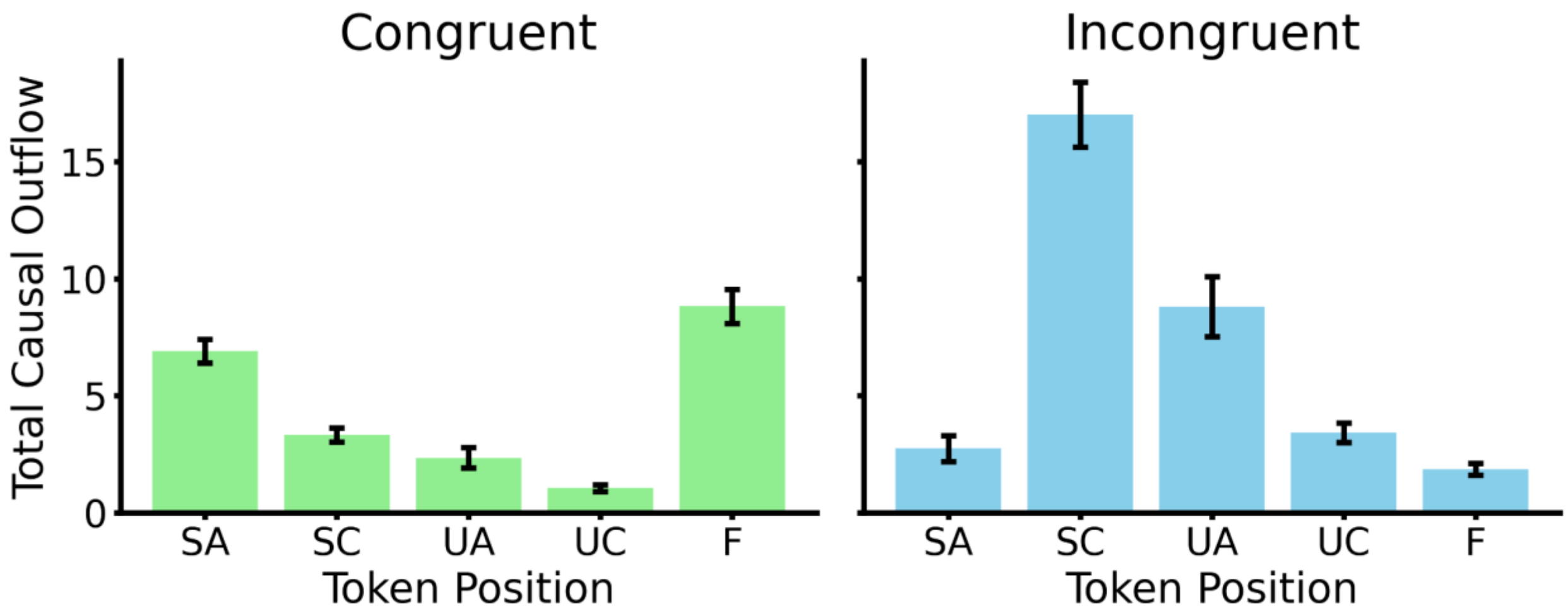

**Figure 3.** *Causal Attribution Outflow by Task Condition*. In the congruent condition, the strongest causal attribution outflow was observed at the final color word position. In the incongruent condition, the strongest causal attribution outflow was observed at the satisfied consequent position. Error bars are 95% confidence intervals. SA=satisfied antecedent; SC= satisfied consequent; UA=unsatisfied antecedent; UC= unsatisfied consequent; F=final color word

**4. There were corresponding condition differences in attention patterns from the final prompt token to the previous color words**

We quantified attention weights from the final token position to the five prompt-associated color word positions in the congruent and incongruent conditions. As shown in Figure 4, left and middle panels, attention heat maps showed clear differences in attention weight patterns across the two conditions, especially at the SC token position. As shown in the right panel of Figure 4, the strongest positive effect (I>C) was at SC (0.0212, t=70.1713, p=1.1587e-42). The strongest negative effect (C>I) was at F (-0.0086, t=-27.5652, p=3.4812e-27).

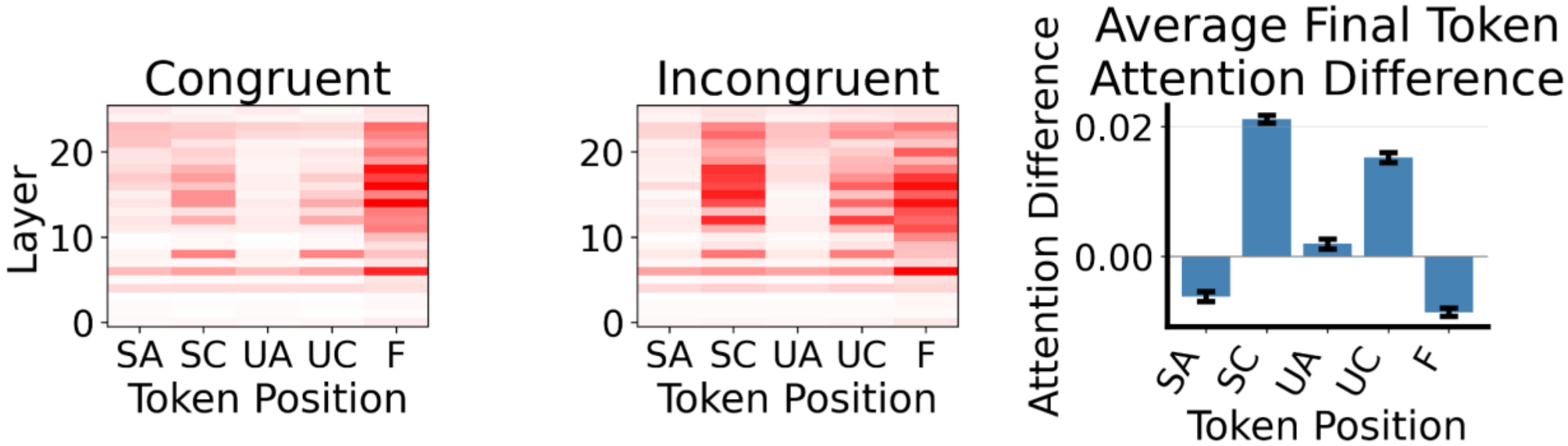

**Figure 4**. *Attention Patterns by Task Condition*. We quantified attention weights from the final token position to the five prompt-associated color word positions. The left and middle panels show layer-wise attention weights separately for the congruent and incongruent conditions. The right panel shows incongruent-congruent difference scores for summed attention weights. Error bars are 95% confidence intervals. SA=satisfied antecedent; SC= satisfied consequent; UA=unsatisfied antecedent; UC= unsatisfied consequent; F=final color word

## 5. Ablating attention to the satisfied consequent disproportionately impairs incongruent condition performance

In the incongruent condition, we observed higher attention weights to, and higher causal attribution outflow from, the SC token position. To further investigate the role of the SC in task performance, we ablated attention weights from all downstream token positions to the SC. As shown in Figure 5, ablation had small impacts in the congruent condition (change in congruent $\Delta$ probability: 0.0962) but a much larger impact in the incongruent condition (change in incongruent $\Delta$ probability: 0.8829), and the change in the incongruent condition was much larger than the change in the congruent condition (incongruent change - congruent change = 0.7864, t=43.1503, p=1.5622e-34).

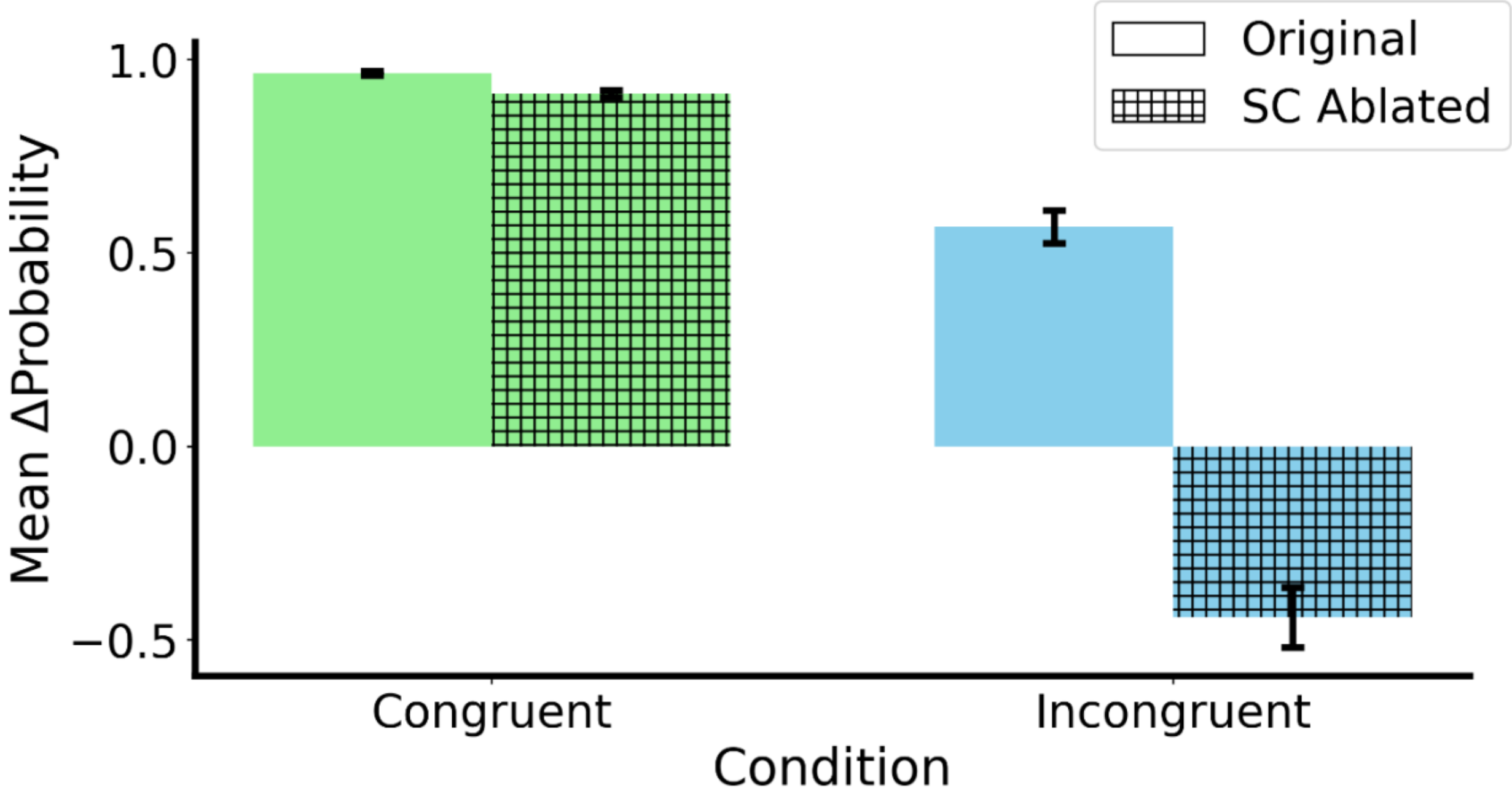

**Figure 5**: *Effects of Attention Ablation by Task Condition*. We ablated attention weights from all downstream token positions to the satisfied consequent (SC). Ablation had small impacts in the congruent condition but a much larger impact in the incongruent condition, and the difference in effects across conditions was highly statistically significant. Error bars are 95% confidence intervals.

**6. Fine tuning the model to systematically enhance the "default" tendency to answer with the same color word yields divergent effects on performance by task condition**

We employed fine tuning to enhance the model's in-weight tendency to produce the same color word response to the stem-only prompt. As shown in Figure 6, this manipulation produced the intended effect starting at roughly the fifth checkpoint. As the tendency to produce the same color word increased, the incongruent *Δ* probability correspondingly dropped (worse performance). The congruent *Δ* probability was already 0.9644 prior to fine tuning, which is likely close to ceiling. It increased to 0.9846 at the final checkpoint of fine tuning (better performance), a small increase that was nonetheless statistically significant ($p=0.001$). in the congruent condition, increased performance (i.e., larger *Δ* probability) at the final checkpoint compared to the pre-fine tuning baseline was also seen across all six Pythia models (410M $p<5.7\times10^{-8}$; 1B $p<4.5\times10^{-6}$; 1.4B $p<1.4\times10^{-6}$; 2.8B $p<0.0002$; 6.9B $p<0.0006$; 12B $p<0.00001$), see Supplement Figure S9. Increases in performance were especially prominent in the first three Pythia models (410M, 1B, 1.4B), in which the congruent *Δ* probability was not already at ceiling prior to fine tuning. Thus, the overall pattern across models reflected divergent effects, in which enhancing a model's "default" tendency to answer with the same color word decreases incongruent condition performance but increases congruent condition performance, an effect that was especially prominent in models that were not already close to ceiling congruent performance.

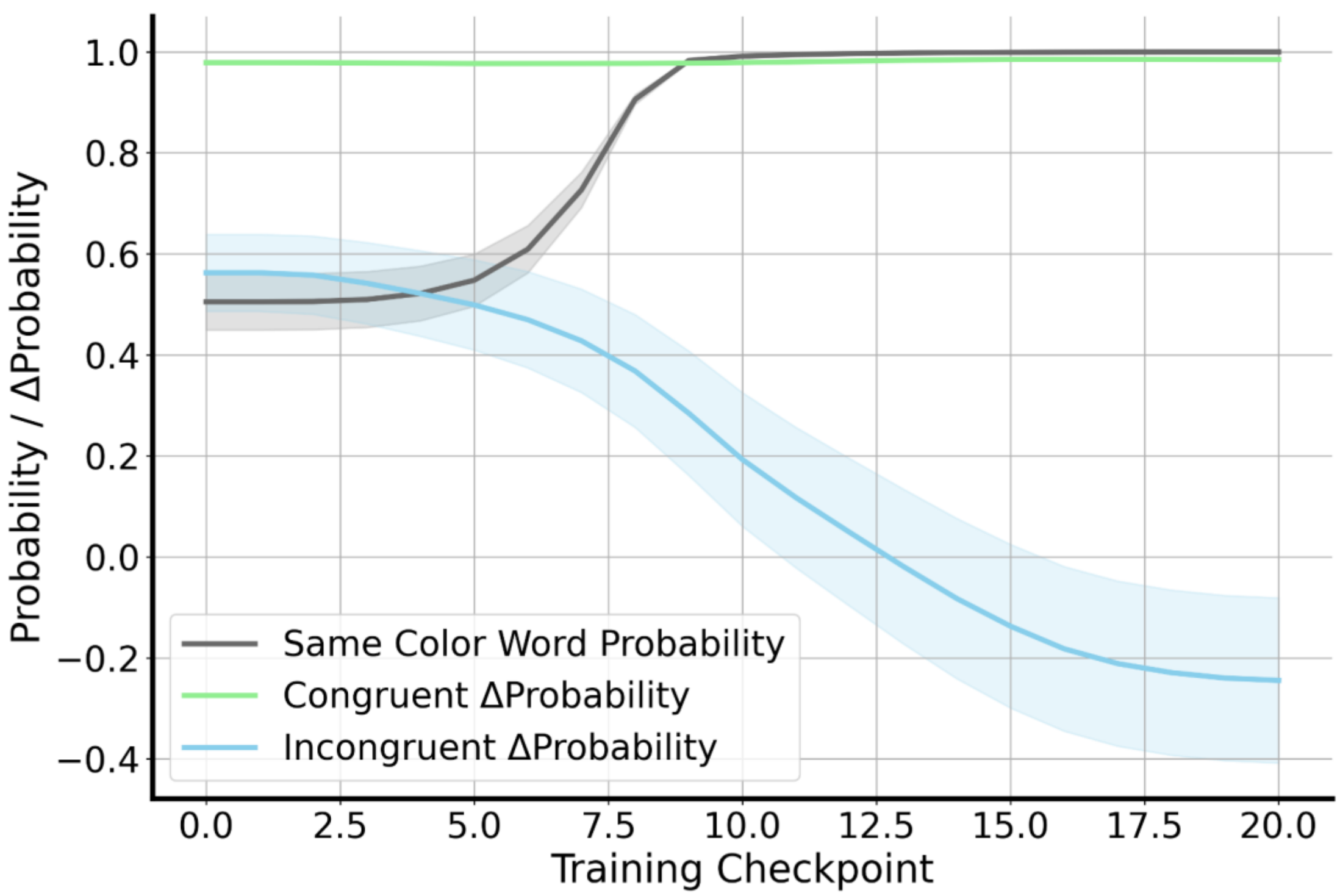

**Figure 6**: *Effects of fine tuning to amplify the default "same color word" tendency in Gemma-2-2B*. Fine tuning produced the intended effect of increasing the model's tendency to produce the same color word response in the stem-only prompt. As the "default" tendency to produce the same color word increased, the incongruent *Δ* probability correspondingly dropped (i.e., worse performance). Shading reflects 95% confidence intervals.

### 7. Increasing rule set size reduces incongruent condition performance (smaller *Δ* probability) but has minimal impact on congruent performance, with effects concentrated on clauses in the middle

We next aimed to manipulate the efficacy of in-context processing. To accomplish this, we increased the number of clauses in the rule-based prefix, systematically assessing model performance (in terms of *Δ* probability) as the number of clauses in the prefix ranged from two to five. As shown in Figure 7, increasing the number of color rule clauses had minimal effects on the congruent condition but had substantial effects on the incongruent condition, with a significant decrease in model performance in the five-clause rule set compared to the original rule with two clauses (decrease in *Δ* probability in five-clause rule compared to two-clause rule = 0.3039, t=3.2171, p=0.002). As shown in Figure 8, the manipulation's effect was strongest when the satisfied consequent was in the middle clauses. For rule-based prefixes with three, four, and five clauses in the incongruent condition, when the satisfied consequent was in the middle clauses, *Δ* probability was statistically significantly reduced compared to the average *Δ* probability when the satisfied consequent was in the first and last clauses (all p value's < 0.001).

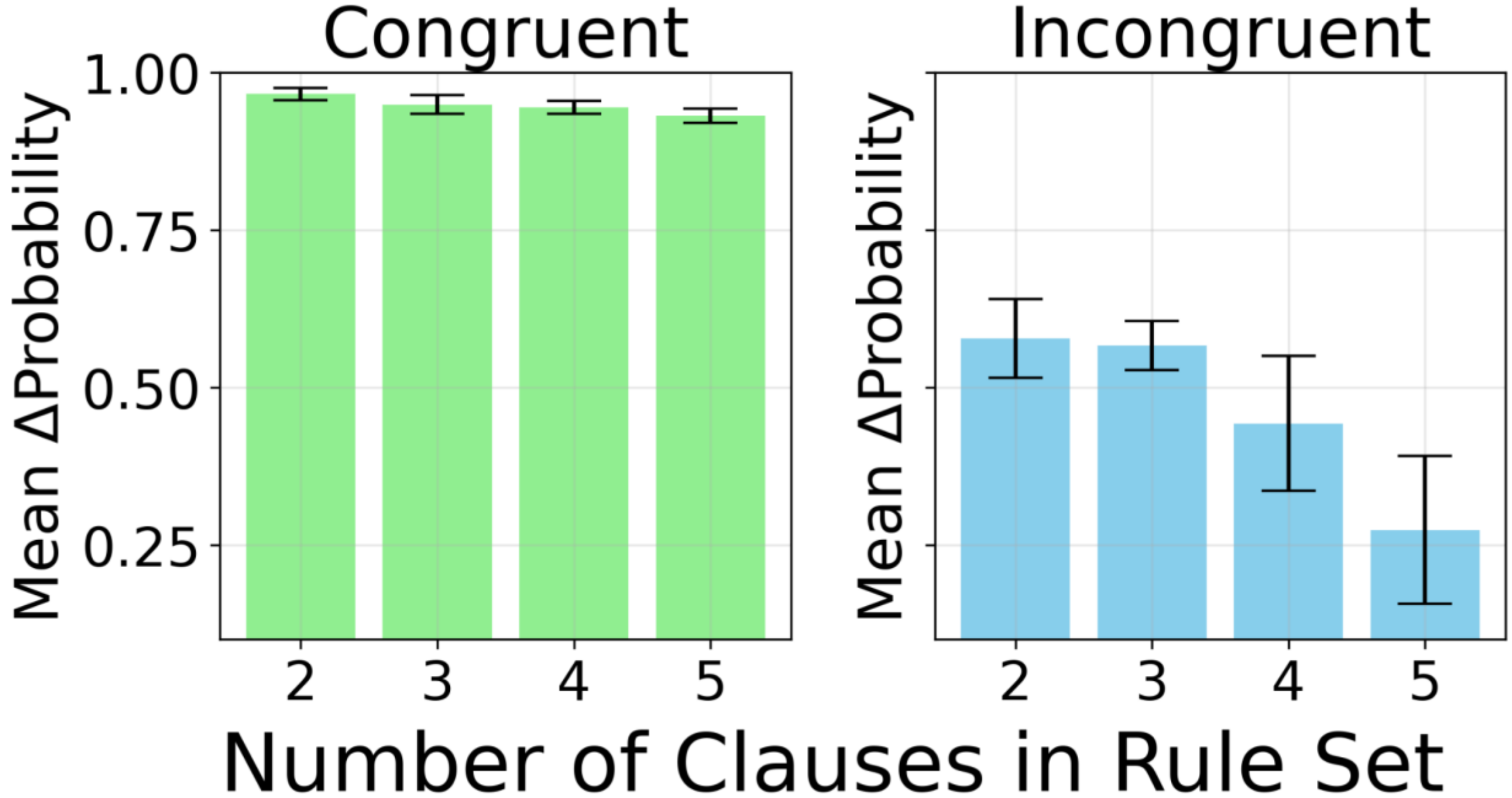

**Figure 7**: *Effects of rule set size manipulation by task condition.* With the aim of selectively manipulating the efficacy of in-context processing, we assessed task performance when the number of color rule clauses ranged from 2 to 5. (Left Panel) Increasing the number of color rule clauses had minimal effects on the congruent condition. (Right Panel) In contrast, it significantly reduced performance (lower *Δ* probability) in the incongruent condition. Error bars are 95% confidence intervals.

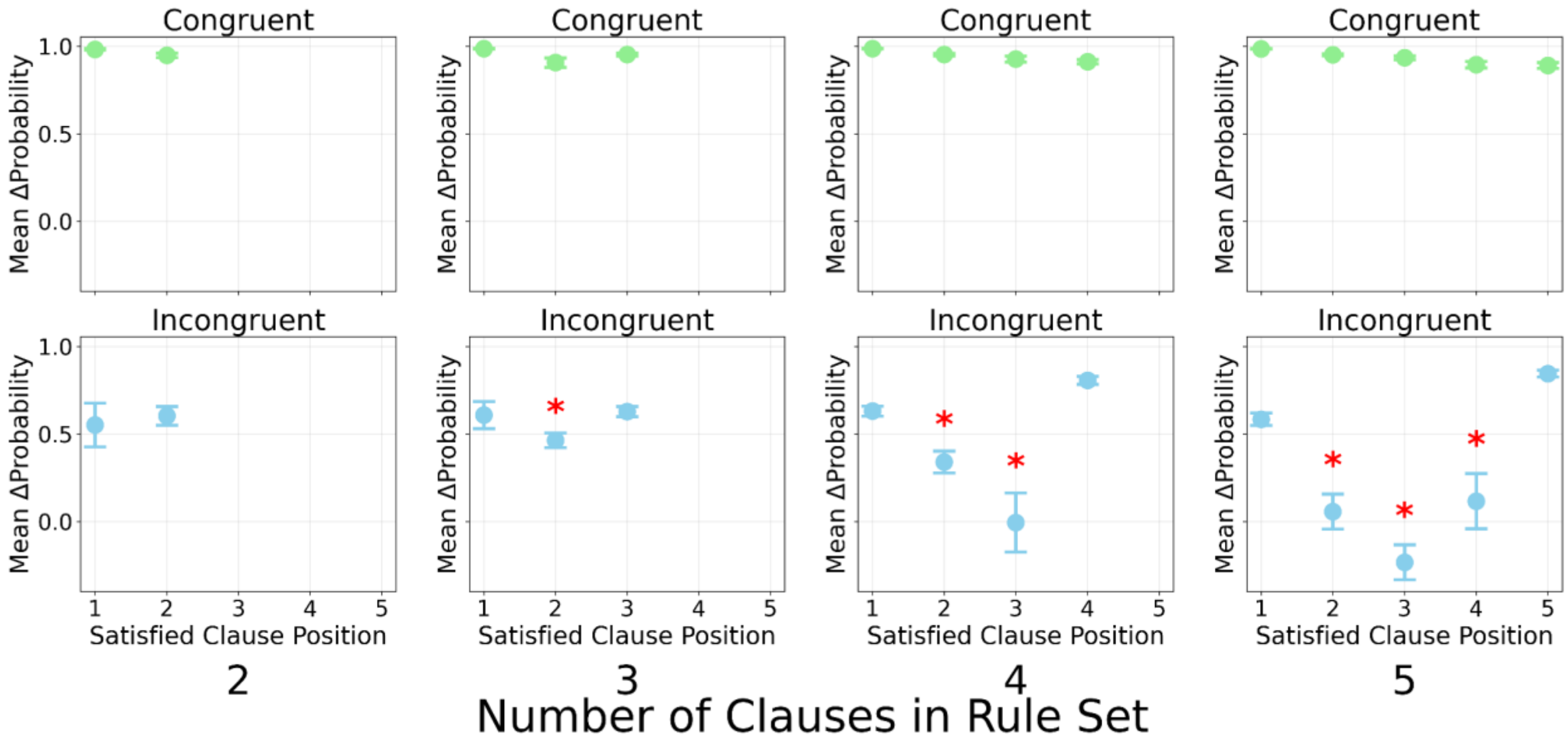

**Figure 8**: *Effects of rule set size manipulation by satisfied clause position.* (Top Panel) In the congruent condition, the rule set size manipulation had minimal impact. (Bottom Panel) In the incongruent condition, the manipulation's impact was strongest when the satisfied consequent was in the middle clauses. Error bars are 95% confidence intervals. * = $p < 0.001$

### 8. Highly similar results were observed across six Pythia models ranging from 410M to 12B

We repeated the preceding analyses on six Pythia models: 410M, 1B, 1.4B, 2.8B, 6.9B, and 12B. As shown in the Supplement, Figures S6 to S10, we found a similar pattern of results across all these models, and they were also highly consistent with the results from Gemma-2-2B. In particular, all these models exhibited a robust congruency effect, except for Pythia-2.8B, in which $\Delta$ probability in congruent and incongruent did not differ (Figure S6). In all the models, the incongruent condition had maximum SC-directed attention, while the congruent condition had maximum F-directed attention (Figure S7), and ablation of attention to SC had greater impact on incongruent than congruent performance (Figure S8). In all the models, fine tuning to enhance the same-color response yielded divergent effects: reduced incongruent $\Delta$ probability and increased congruent $\Delta$ probability (Figure S9). In contrast, increasing rule set size reduced incongruent $\Delta$ probability, specifically at the middle clause positions, while leaving congruent $\Delta$ probability unaffected (Figure S10).

## Discussion

In this study, we introduced a novel verbal-only conflict task and examined congruency effects in Gemma-2-2B as well as a series of Pythia models. We applied methods from mechanistic interpretability—to the best of our knowledge, for the first time—to understand internal processing patterns that help explain the generation of these congruency effects. We identified distinct processing pathways that are preferentially activated under congruent versus incongruent conditions. The overall pattern of results, including from fine-tuning and rule set size manipulations, supports competition between in-weight and in-context processing as key contributors to congruency effects in this task. More broadly, this study introduces new approaches for investigating internal processing patterns in LLMs during conflict tasks and underscores the role of capable LLMs as artificial model systems to help understand complex cognitive phenomena.

Multiple lines of evidence point to the presence of separable internal processing pathways in the LLMs that help to explain task performance. We found that the stem-alone prompt produces a robust “default” tendency for a same-color completion. But when the rule-based prefix was appended to the stem, we observed robust congruency effects across nearly all the models in which response evidence was lower in the incongruent condition than in the congruent condition. A plausible explanation for this effect, analogous to a standard view of Stroop task performance (MacLeod 1991), is that the default response tendency conflicts with the rule-based mapping in the incongruent condition, resulting in lower overall strength of evidence for the correct response. However, it aligns with the rule-based mapping in the congruent condition, resulting in higher overall strength of evidence for the correct response.

Additional support for separable internal processing pathways comes from causal attribution graphs, which use ablations to estimate feature-to-feature direct causal effects (Lindsey et al. 2025). This method identified distinct causal geometries associated with the congruent and incongruent conditions. In the congruent condition, the strongest causal attribution outflow to the correct color word response emerged from the final color word, a local, easily accessed cue in the prompt stem. In the incongruent condition, in contrast, the final color word is a misleading cue. Instead, the strongest causal attribution outflow to the correct color word response emerged from the satisfied consequent position, indicating significantly greater impact of the rule-based prefix—a more distant cue that requires interpretation of a logical conditional to properly utilize.

Distinct causal geometries in the two conditions were mirrored and potentially explained in terms of changes in attentional patterns. In the congruent condition, attention from the final token position was highest for the final color word, while in the incongruent condition, it was highest for the satisfied consequent. When attention to the satisfied consequent was ablated, model performance collapsed specifically for the incongruent condition but was largely preserved for the congruent condition. This result reinforces the idea that the model requires the rule-based prefix to generate the correct response only in the incongruent condition. In the congruent condition, the model can rely on an alternative pathway involving the final color word, and thus ablating attention to the satisfied consequent has only minimal impact on performance.

Taken together, the preceding observations are consistent with an interpretation in which performance on the crayon conflict task, and specifically congruency effects in this task, are explained in terms of competition between in-weight processing and in-context processing. Further supporting this interpretation, we observed strong same-color word tendencies in the prompt stem alone, which is plausibly attributable to an in-weight tendency learned during model training to say that something has a feature when it has been described as having that feature—for example, in response to the prompt “The crayon is blue, so you say the crayon is”, the model responds with *blue*. In contrast, the rule-based prefix involves a novel antecedent–consequent logical mapping specified within the prompt that must be interpreted and applied online, a canonical form of in-context task learning (T. Brown et al. 2020; Pan et al. 2023; Bhattamishra et al. 2024).

To provide an additional direct test of the role of in-weight/in-context competition in producing congruency effects in the crayon task, we performed a pair of manipulations that plausibly have selective impacts on each type of processing. We used fine-tuning to selectively augment in-weight tendencies to produce the same-color response. This manipulation is predicted to produce divergent effects across the two conditions: It should decrease performance (i.e., lower $\Delta$ probability) in the incongruent condition, since the in-weight tendency to produce the same-color response impedes the correct response in this condition. But it should increase performance (i.e., higher $\Delta$ probability) in the congruent condition, since the in-weight tendency to produce the same-color response facilitates the correct response in this condition, and this divergent pattern was in fact observed.

In contrast, increasing the number of clauses in the rule-based mapping should selectively degrade in-context processing, which is implicated in the processing of novel rules (T. Brown et al. 2020; Bhattamishra et al. 2024). It is thus predicted to reduce incongruent performance (i.e., lower $\Delta$ probability) but have little impact on congruent performance, which is the pattern we in fact observed. More specifically, we observed decreases in performance concentrated in the middle clauses of the rule-based prefix. This pattern is predicted based on the “lost in the middle” effect observed in transformers in which the first and last items in a series have a processing advantage (Liu et al. 2024). It is also notable that an analogous effect is observed in studies of human list memory (Ji-An et al. 2024).

Importantly, the in-weight/in-context processing distinction is likely not unique to LLMs and a number of studies have applied it to illuminate aspects of human cognition (Russin et al. 2025; Pesnot Lerousseau and Summerfield 2026; Ji-An et al. 2024; Schubert et al. 2024). The present study extends this approach to help understand human conflict task performance and the etiology of congruency effects.

It is notable that the in-weight/in-context processing distinction has strong face plausibility as corresponding to the distinction between automatic and controlled processing (Schneider and Shiffrin 1977; Shiffrin and Schneider 1977), which has been highly influential in explaining conflict task performance (MacLeod 1991; Cohen et al. 1990) and, more broadly, in theories of reasoning, judgment, and social cognition (Evans 2008). Similar to in-weight processing, automatic processing is often characterized as dependent on extensive prior training and reliant on shallow stimulus-based cues (Shiffrin and Schneider 1977; Moors and De Houwer 2006). In the present study, we found that the congruent condition was associated with preferential engagement of short-range attention to the final color word (arguably, analogous to a shallow, stimulus-based cue). We also found that additional training through fine-tuning, which plausibly increases in-weight response tendencies, yielded divergent effects: it decreased incongruent condition performance but increased congruent condition performance. It is notable that in conflict tasks such as the Stroop task, automatic processing is also thought to contribute to this divergent pattern. According to an influential model, automatic processing opposes task-associated controlled processing in the incongruent condition, decreasing task performance (the Stroop interference effect). But automatic processing aligns with controlled processing in the congruent condition, increasing task performance and producing the so-called Stroop facilitation effect (Dalrymple-Alford 1972; MacLeod 1991; Roelofs 2010).

On the other hand, similar to in-context processing, controlled processing is characterized as being flexibly responsive to task rules and sensitive to cognitive load (Shiffrin and Schneider 1977). In the present study, we found performance in the incongruent condition preferentially relies on an in-context processing pathway that involves long-range attention to the satisfied consequent in the rule-based prefix, and ablating attention to the satisfied consequent impaired the ability of the model to respond to the rule set. The role of this pathway in flexible rule following suggests it plays a role analogous to controlled processing in theories of human task performance. Additionally, we showed selective effects of rule set size on in-context processing in the incongruent condition. This selective impairment of incongruent responding parallels a recurrent finding in human cognition: working-memory or executive load disproportionately impairs performance when controlled processing must override a competing automatic response (Lavie et al. 2004; Neys 2006; Lavie 2010; Pratt et al. 2011).

Taken together, these observations suggest the in-weight/in-context distinction in LLMs may be closely related to the automatic/controlled distinction in cognitive science, a possibility that warrants deeper, systematic exploration. More broadly, these results underscore a potential role for LLMs to serve as artificial model systems for complex human capacities, where mechanisms uncovered in LLMs can help illuminate analogous mechanisms operative in human cognition.

Some theorists explain conflict task performance, and the phenomena of automaticity and control more broadly, in terms of competition between distinct mentally instantiated processors that operate with different processing principles (e.g., an executive processor that is rule-governed and symbolic vs. an automatic processor that is associative) (Posner et al. 2004; Shiffrin and Schneider 1977; Sloman 1996; Evans 2008; Kahneman 2011). Other theorists appeal to distinct processing pathways within a single network, for example the PDP model in Cohen, Dunbar, and McClelland (1990) discussed earlier. Results from this study suggest a middle position that incorporates elements of both. We find congruency effects and signatures of competing processes in a number of LLMs, despite the fact that these models are single, "unified" networks that lack modular, specialized subprocessors. Instead, what conflicts in these LLMs are two *pathways* or *trajectories* through the transformer, one involving local attention to easily accessible cues and one involving long-range attention to the rule-based color mappings. Importantly, however, the two pathways are not equivalent; our results suggest they implicate distinct types of processing, in particular, in-weight versus in-context processing (Radford et al. 2019; T. Brown et al. 2020). Thus, LLMs illustrate a middle position in which a single network without distinct, modular *processors* can nevertheless give rise to qualitatively different *processes*, and indeed, these processes can antagonistically compete in the way envisioned in traditional dual-process theories.

We replicated our analyses across a series of Pythia models of increasing model size (with the exception of causal attribution analysis, which is not available for Pythia models). In all six Pythia models, we observed a broadly similar pattern of results: stem-alone prompts produce "default" dispositions to respond with the same color word; congruency effects, with one exception to be discussed; distinct attention patterns from the final token in the congruent condition (final color word strongest) compared to the incongruent condition (satisfied consequent strongest); attention ablation to the SC position differentially harms incongruent performance; divergent effects across conditions from fine tuning; and selective effects of increased rule set size on incongruent responding. These findings point to the generality of our explanation of conflict task performance in terms of in-context/in-weight competition across models that vary in size.

A notable exception was Pythia-2.8B, which failed to show a congruency effect (i.e., congruent and incongruent *Δ* probability were not different). We do not, however, interpret this null finding as evidence that the underlying competition between in-context and in-weight processing was absent. Because in-context and in-weight influences may combine nonlinearly or may be moderated by other factors such as learned attention routing profiles, the resulting interference effect may exhibit idiosyncratic patterns (Longpre et al. 2021; Zhao et al. 2026). Of note, in the Pythia-2.8B model, we did observe differential effects across congruent and incongruent conditions for all the other assessed effects (e.g., effects of attention ablation, fine-tuning manipulation, rule set size manipulation, etc.). Thus, there is evidence consistent with the Pythia-2.8B model being sensitive to congruent/incongruent differences, suggesting that in-context/in-weight competition may have been present in the model, even if a congruency effect in *Δ* probability and *Δ* log probability was absent.

This study has several limitations and invites a number of future directions for research. First, we used a verbal conflict task in order to maximize interpretability of model processing, given limitations in the existing mechanistic interpretability toolkit for multimodal models such as vision language models. As interpretability tools for multimodal models mature and become further validated, future work should seek to corroborate our findings using mechanistic interpretability tools applied to standard conflict tasks such as the Stroop or flanker tasks. Second, this study focused specifically on the congruency effect, one of the best-known and most extensively studied phenomena in the conflict task literature (Stroop 1935; MacLeod 1991). Future studies can build on the present work to examine a wider range of conflict task phenomena, including congruency sequence effects (Gratton et al. 1992; Duthoo et al. 2014), post-error adjustments (Rabbitt 1966; Danielmeier and Ullsperger 2011), and effects of changes in performance as a function of the relative frequency of congruent and incongruent trials (Logan and Zbrodoff 1979; Bugg and Crump 2012). The mechanistic bases of these effects remain poorly understood, with theorists putting forward a number of competing hypotheses to explain each one. Future work can leverage LLMs as model systems to test competing hypotheses and as a source of new theoretical frameworks to better understand the source of these effects.

In sum, this study is, to the best of our knowledge, the first to systematically apply mechanistic interpretability tools to uncover processing patterns that help explain congruency effects in a novel verbal conflict task in LLMs. Our results point to competition between in-weight and in-context processing as a critical mechanism of the congruency effect in this conflict task, inviting future work in human cognitive neuroscience that adopts this theoretical framing to better understand the mechanistic underpinnings of conflict task performance.

# References

Ameisen, Emmanuel, Jack Lindsey, Adam Pearce, et al. 2025. “Circuit Tracing: Revealing Computational Graphs in Language Models.” *Transformer Circuits Thread* 6 (16318–16352): 1.

Bari, Andrea, and Trevor W. Robbins. 2013. “Inhibition and Impulsivity: Behavioral and Neural Basis of Response Control.” *Progress in Neurobiology* 108: 44–79.

Barkley, Russell A. 1997. “Behavioral Inhibition, Sustained Attention, and Executive Functions: Constructing a Unifying Theory of ADHD.” *Psychological Bulletin* 121 (1): 65–94. https://doi.org/10.1037/0033-2909.121.1.65.

Bhattamishra, Satwik, Arkil Patel, Phil Blunsom, and Varun Kanade. 2024. “Understanding In-Context Learning in Transformers and Llms by Learning to Learn Discrete Functions.” *International Conference on Learning Representations* 2024: 56094–126.

Biderman, Stella, Hailey Schoelkopf, Quentin Gregory Anthony, et al. 2023. “Pythia: A Suite for Analyzing Large Language Models across Training and Scaling.” *International Conference on Machine Learning*, 2397–430.

Binz, Marcel, and Eric Schulz. 2023. “Using Cognitive Psychology to Understand GPT-3.” *Proceedings of the National Academy of Sciences* 120 (6): e2218523120.

Botvinick, Matthew M., T. S. Braver, D. M. Barch, C. S. Carter, and J. D. Cohen. 2001. “Conflict Monitoring and Cognitive Control.” *Psychol Rev* 108: 624–52.

Brown, Scott D., and Andrew Heathcote. 2008. “The Simplest Complete Model of Choice Response Time: Linear Ballistic Accumulation.” *Cognitive Psychology* 57 (3): 153–78.

Brown, Tom, Benjamin Mann, Nick Ryder, et al. 2020. “Language Models Are Few-Shot Learners.” *Advances in Neural Information Processing Systems* 33: 1877–901.

Bugg, Julie M., and Matthew JC Crump. 2012. “In Support of a Distinction between Voluntary and Stimulus-Driven Control: A Review of the Literature on Proportion Congruent Effects.” *Frontiers in Psychology* 3: 367.

Cohen, J. D., K. Dunbar, and J. L. McClelland. 1990. “On the Control of Automatic Processes: A Parallel Distributed Processing Account of the Stroop Effect.” *Psychological Review* 97 (3): 332–61.

Dalrymple-Alford, E. C. 1972. “Associative Facilitation and Interference in the Stroop Color-Word Task.” *Perception & Psychophysics* 11 (4): 274–76.

Danielmeier, Claudia, and Markus Ullsperger. 2011. “Post-Error Adjustments.” *Frontiers in Psychology* 2: 233.

Duthoo, Wout, Elger L. Abrahamse, Senne Braem, Carsten N. Boehler, and Wim Notebaert. 2014. “The Heterogeneous World of Congruency Sequence Effects: An Update.” *Frontiers in Psychology* 5: 1001.

Elhage, Nelson, Neel Nanda, Catherine Olsson, et al. 2021. “A Mathematical Framework for Transformer Circuits.” *Transformer Circuits Thread* 1 (1): 12.

Eriksen, Barbara A., and Charles W. Eriksen. 1974. “Effects of Noise Letters upon the Identification of a Target Letter in a Nonsearch Task.” *Perception & Psychophysics* 16 (1): 143–49.

Evans, Jonathan St B. T. 2008. “Dual-Processing Accounts of Reasoning, Judgment, and Social Cognition.” *Annual Review of Psychology* 59: 255–78. https://doi.org/10.1146/annurev.psych.59.103006.093629.

Gemma Team, Morgane Riviere, Shreya Pathak, et al. 2024. “Gemma 2: Improving Open Language Models at a Practical Size.” *arXiv Preprint arXiv:2408.00118*.

Goldstein, Rita Z., and Nora D. Volkow. 2011. “Dysfunction of the Prefrontal Cortex in Addiction: Neuroimaging Findings and Clinical Implications.” *Nature Reviews Neuroscience* 12 (11): 652–69.

Gratton, Gabriele, Michael GH Coles, and Emanuel Donchin. 1992. “Optimizing the Use of Information: Strategic Control of Activation of Responses.” *Journal of Experimental Psychology: General* 121 (4): 480.

Hanna, Michael, Mateusz Piotrowski, Jack Lindsey, and Emmanuel Ameisen. 2025. “Circuit-Tracer: A New Library for Finding Feature Circuits.” *Proceedings of the 8th BlackboxNLP Workshop: Analyzing and Interpreting Neural Networks for NLP*, 239–49.

Herd, Seth A., Marie T. Banich, and Randall C. O’reilly. 2006. “Neural Mechanisms of Cognitive Control: An Integrative Model of Stroop Task Performance and fMRI Data.” *Journal of Cognitive Neuroscience* 18 (1): 22–32.

Hu, Edward J., Yelong Shen, Phillip Wallis, et al. 2021. “Lora: Low-Rank Adaptation of Large Language Models.” *arXiv Preprint arXiv:2106.09685*.

Hu, Jennifer, Michael A. Lepori, and Michael Franke. 2025. “Signatures of Human-like Processing in Transformer Forward Passes.” *arXiv Preprint arXiv:2504.14107*.

Hu, Xiaoyang. 2025. “Conflict Adaptation in Vision-Language Models.” *arXiv Preprint arXiv:2510.24804*.

Ji-An, Li, Corey Y. Zhou, Marcus K. Benna, and Marcelo G. Mattar. 2024. “Linking In-Context Learning in Transformers to Human Episodic Memory.” *Advances in Neural Information Processing Systems* 37: 6180–212.

Jones, Cameron R., Sean Trott, and Benjamin Bergen. 2024. “Comparing Humans and Large Language Models on an Experimental Protocol Inventory for Theory of Mind Evaluation (EPITOME).” *Transactions of the Association for Computational Linguistics* 12: 803–19.

Kahneman, Daniel. 2011. *Thinking, Fast and Slow*. 1st ed. Farrar, Straus and Giroux.

Kalanthroff, Eyal, Eddy J. Davelaar, Avishai Henik, Liat Goldfarb, and Marius Usher. 2018. "Task Conflict and Proactive Control: A Computational Theory of the Stroop Task." *Psychological Review* 125 (1): 59.

Lampinen, Andrew K., Ishita Dasgupta, Stephanie CY Chan, et al. 2024. "Language Models, like Humans, Show Content Effects on Reasoning Tasks." *PNAS Nexus* 3 (7): pgae233.

Lavie, Nilli. 2010. "Attention, Distraction, and Cognitive Control under Load." *Current Directions in Psychological Science* 19 (3): 143–48.

Lavie, Nilli, Aleksandra Hirst, Jan W. De Fockert, and Essi Viding. 2004. "Load Theory of Selective Attention and Cognitive Control." *Journal of Experimental Psychology: General* 133 (3): 339.

Lieberum, Tom, Senthooran Rajamanoharan, Arthur Conmy, et al. 2024. "Gemma Scope: Open Sparse Autoencoders Everywhere All at Once on Gemma 2." *Proceedings of the 7th BlackboxNLP Workshop: Analyzing and Interpreting Neural Networks for NLP*, 278–300.

Lindsey, Jack, Wes Gurnee, Emmanuel Ameisen, et al. 2025. "On the Biology of a Large Language Model." *Transformer Circuits Thread*.

Liu, Nelson F., Kevin Lin, John Hewitt, et al. 2024. "Lost in the Middle: How Language Models Use Long Contexts." *Transactions of the Association for Computational Linguistics* 12: 157–73.

Logan, Gordon D., and N. Jane Zbrodoff. 1979. "When It Helps to Be Misled: Facilitative Effects of Increasing the Frequency of Conflicting Stimuli in a Stroop-like Task." *Memory & Cognition* 7 (3): 166–74.

Longpre, Shayne, Kartik Perisetla, Anthony Chen, Nikhil Ramesh, Chris DuBois, and Sameer Singh. 2021. "Entity-Based Knowledge Conflicts in Question Answering." *Proceedings of the 2021 Conference on Empirical Methods in Natural Language Processing*, 7052–63.

Luo, Dezhi, Maijunxian Wang, Bingyang Wang, Tianwei Zhao, Yijiang Li, and Hokin Deng. 2025. *Machine Psychophysics: Cognitive Control in Vision-Language Models*.

MacLeod, Colin M. 1991. "Half a Century of Research on the Stroop Effect: An Integrative Review." *Psychological Bulletin* 109: 163–203.

Mangrulkar, Sourab, Sylvain Gugger, Lysandre Debut, Younes Belkada, Sayak Paul, and Benjamin Bossan. 2022. *Peft: State-of-the-Art Parameter-Efficient Fine-Tuning Methods*.

Miller, E. K., and J. D. Cohen. 2001. "An Integrative Theory of Prefrontal Cortex Function." *Annu Rev Neurosci* 24: 167–202.

Moors, Agnes, and Jan De Houwer. 2006. "Automaticity: A Theoretical and Conceptual Analysis." *Psychological Bulletin* 132 (2): 297.

Neys, Wim De. 2006. "Dual Processing in Reasoning: Two Systems but One Reasoner." *Psychological Science* 17 (5): 428–33.

Olah, Chris, Nick Cammarata, Ludwig Schubert, Gabriel Goh, Michael Petrov, and Shan Carter. 2020. “Zoom in: An Introduction to Circuits.” *Distill* 5 (3): e00024-001.

Olsson, Catherine, Nelson Elhage, Neel Nanda, et al. 2022. “In-Context Learning and Induction Heads.” *arXiv Preprint arXiv:2209.11895*.

Pan, Jane, Tianyu Gao, Howard Chen, and Danqi Chen. 2023. “What In-Context Learning ‘Learns’ in-Context: Disentangling Task Recognition and Task Learning.” *Findings of the Association for Computational Linguistics: ACL 2023*, 8298–319.

Pesnot Lerousseau, Jacques, and Christopher Summerfield. 2026. “Shared Sensitivity to Data Distribution during Learning in Humans and Transformer Networks.” *Nature Human Behaviour* 10 (3): 601–14.

Posner, Michael I., Charles R. Snyder, and Robert Solso. 2004. “Attention and Cognitive Control.” *Cognitive Psychology: Key Readings* 205: 55–85.

Pratt, Nikki, Adrian Willoughby, and Diane Swick. 2011. “Effects of Working Memory Load on Visual Selective Attention: Behavioral and Electrophysiological Evidence.” *Frontiers in Human Neuroscience* 5: 57.

Rabbitt, PM A. 1966. “Errors and Error Correction in Choice-Response Tasks.” *Journal of Experimental Psychology* 71 (2): 264.

Radford, Alec, Jeffrey Wu, Rewon Child, David Luan, Dario Amodei, and Ilya Sutskever. 2019. “Language Models Are Unsupervised Multitask Learners.” *OpenAI Blog* 1 (8): 9.

Ratcliff, Roger, and Jeffrey N. Rouder. 1998. “Modeling Response Times for Two-Choice Decisions.” *Psychological Science* 9 (5): 347–56.

Roelofs, Ardi. 2010. “Attention and Facilitation: Converging Information versus Inadvertent Reading in Stroop Task Performance.” *Journal of Experimental Psychology: Learning, Memory, and Cognition* 36 (2): 411.

Russin, Jacob, Ellie Pavlick, and Michael J. Frank. 2025. “Parallel Trade-Offs in Human Cognition and Neural Networks: The Dynamic Interplay between in-Context and in-Weight Learning.” *Proceedings of the National Academy of Sciences* 122 (35): e2510270122.

Schneider, Walter, and Richard M. Shiffrin. 1977. “Controlled and Automatic Human Information Processing: I. Detection, Search, and Attention.” *Psychological Review* 84 (1): 1.

Schubert, Johannes A., Akshay K. Jagadish, Marcel Binz, and Eric Schulz. 2024. “In-Context Learning Agents Are Asymmetric Belief Updaters.” *arXiv Preprint arXiv:2402.03969*.

Shiffrin, Richard M., and Walter Schneider. 1977. “Controlled and Automatic Human Information Processing: II. Perceptual Learning, Automatic Attending and a General Theory.” *Psychological Review* 84 (2): 127.

Simon, J. Richard. 1969. “Reactions toward the Source of Stimulation.” *Journal of Experimental Psychology* 81 (1): 174.

Sloman, S. A. 1996. “The Empirical Case for Two Systems of Reasoning.” *Psychological Bulletin* 119: 2–22.

Stroop, J. Ridley. 1935. “Studies of Interference in Serial Verbal Reactions.” *Journal of Experimental Psychology* 18 (6): 643.

Wang, Kevin, Alexandre Variengien, Arthur Conmy, Buck Shlegeris, and Jacob Steinhardt. 2022. “Interpretability in the Wild: A Circuit for Indirect Object Identification in GPT-2 Small.” *arXiv Preprint arXiv:2211.00593*.

Zhao, Jun, Yongzhuo Yang, Xiang Hu, et al. 2026. “Understanding Parametric and Contextual Knowledge Reconciliation within Large Language Models.” *Advances in Neural Information Processing Systems* 38: 102978–3012.

# Supplementary Results

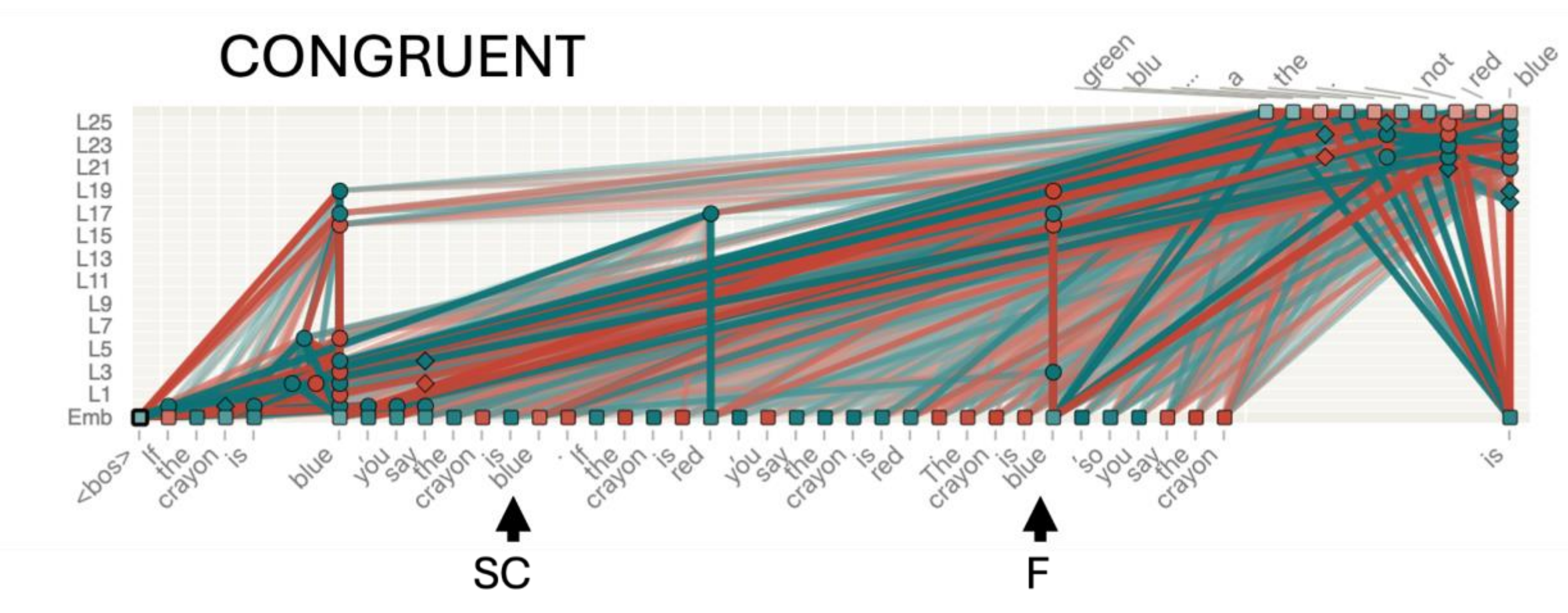

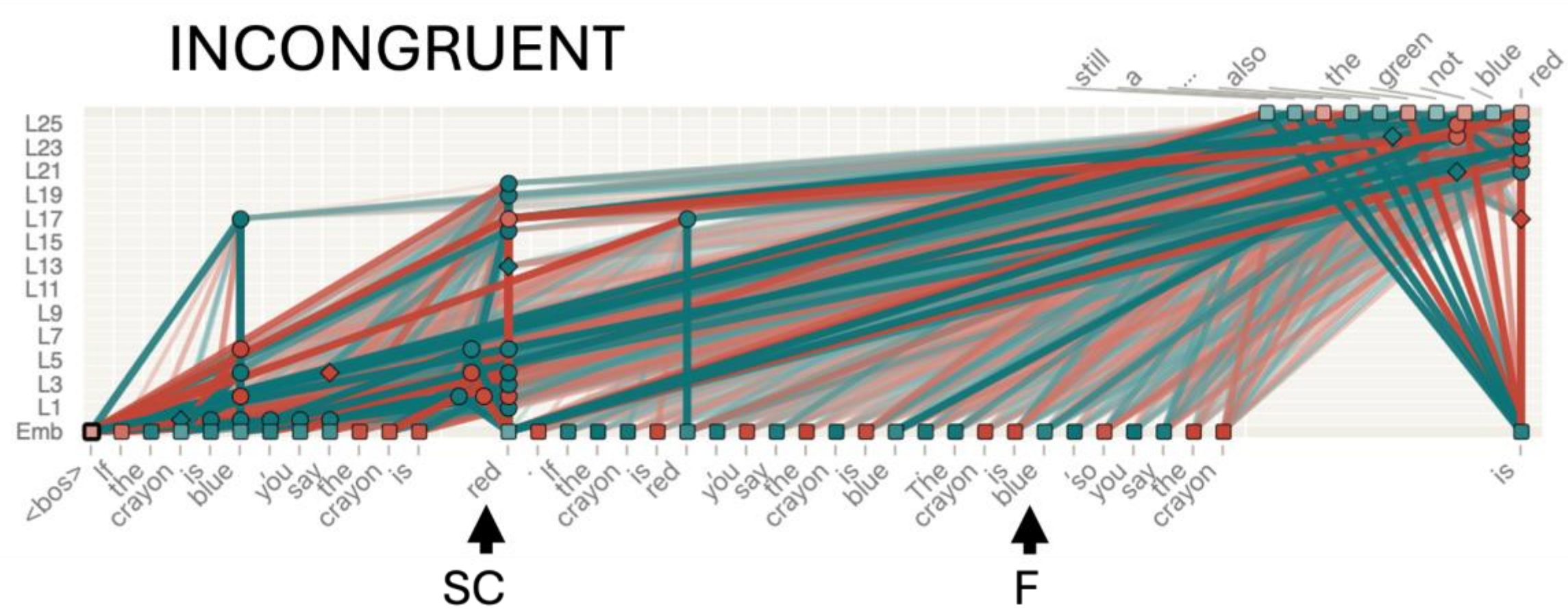

**Figure S1**: *Example Causal Attribution Graphs for Congruent and Incongruent Prompts*. Nodes represent prompt tokens, activated features, and output tokens, and weighted directed edges estimate how strongly each node promotes or suppresses downstream features and, ultimately, the target output. In the congruent condition (top panel), there are more activated features and edges at the final color word position (F), while in the incongruent condition (bottom panel), there are more activated features and edges at the satisfied consequent position. For visualization purposes, graphs are thresholded to show only the top 50% of activated features and edges. Note that nodes and edges are colored strictly to enhance discernibility of overlapping elements in the figure.

Log Probability Results for Gemma-2-2B

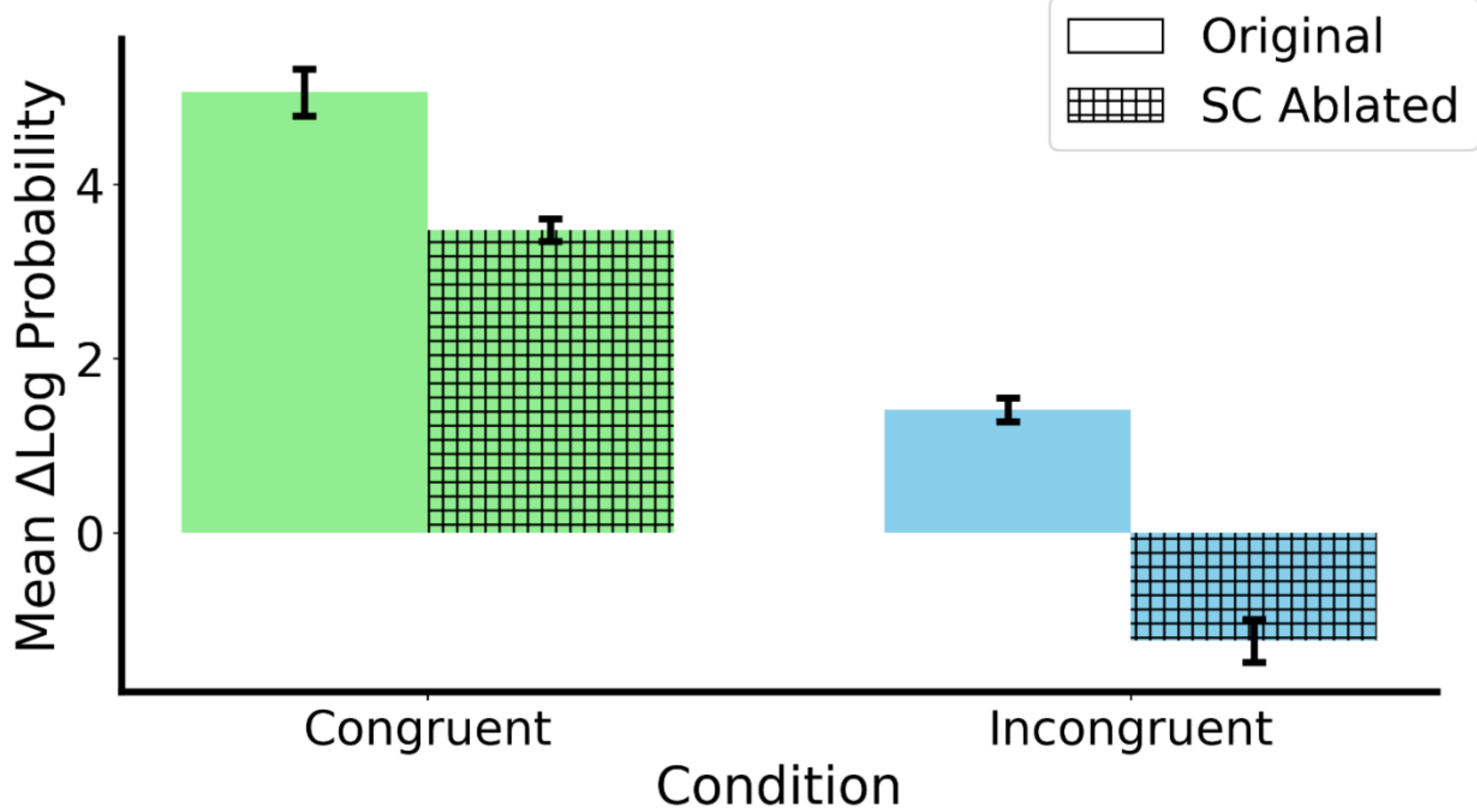

**Figure S2**: *Effects of Attention Ablation by Task Condition*.

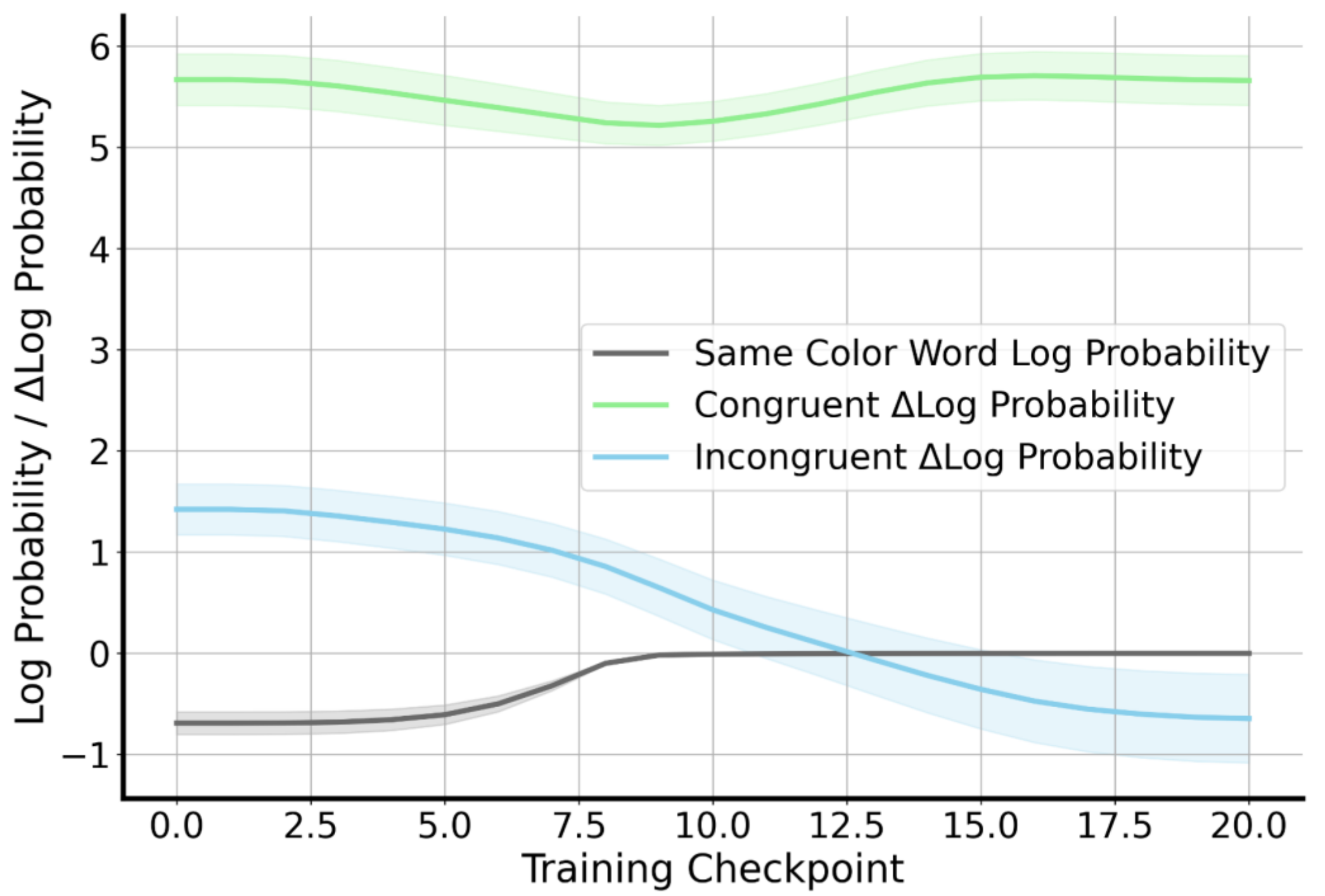

**Figure S3**: *Effects of fine tuning to amplify the default "same color word" tendency in Gemma-2-2B*.

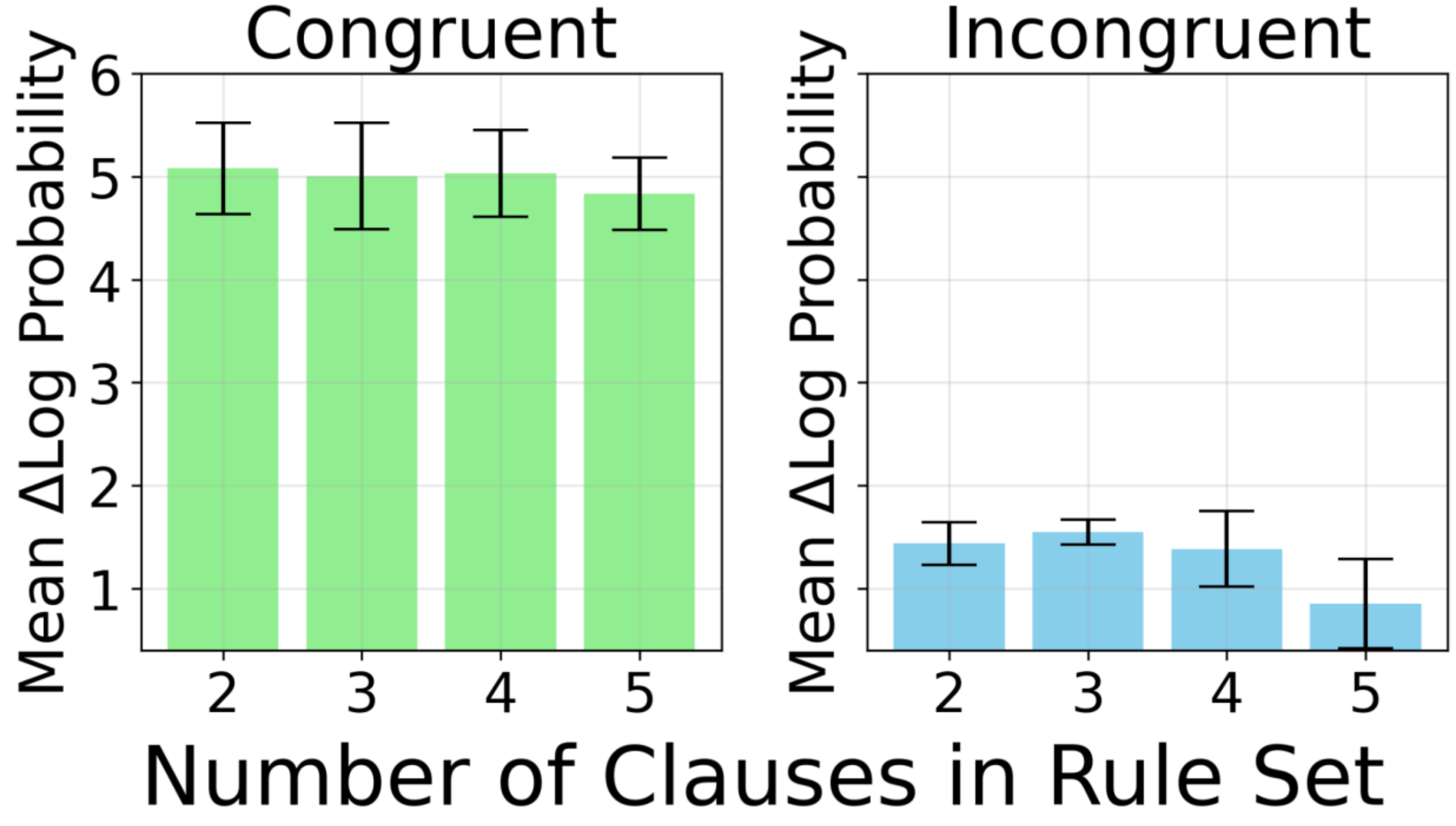

**Figure S4**: *Effects of rule set size manipulation by task condition.*

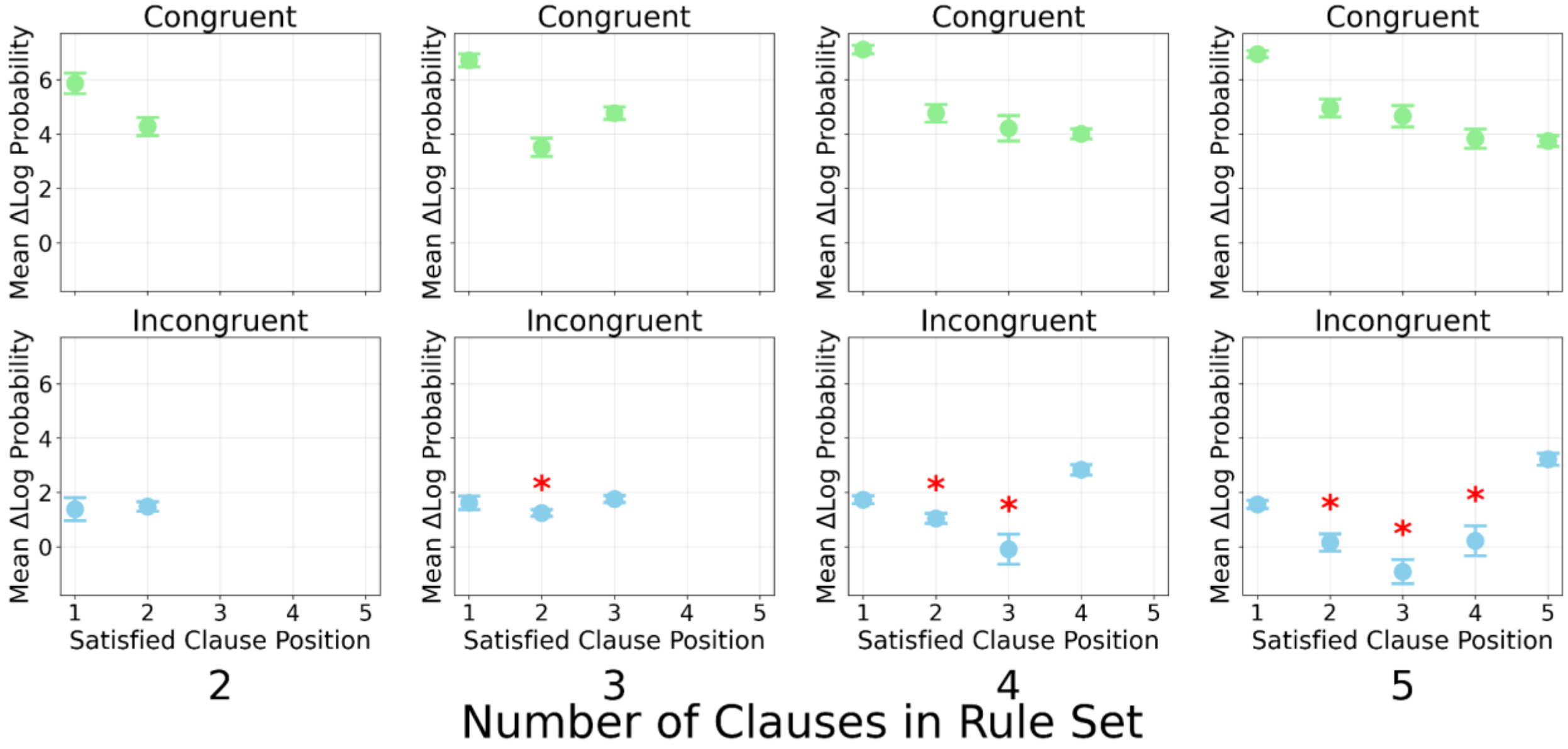

**Figure S5**: *Effects of rule set size manipulation by satisfied clause position.*

## Pythia Results

The analyses reported in the Main Manuscript were performed on Gemma-2-2B. We repeated all analyses on six Pythia models, 410M, 1B, 1.4B, 2.8B, 6.9B, 12B, with the exception of causal attribution analyses which are not available on Pythia models.

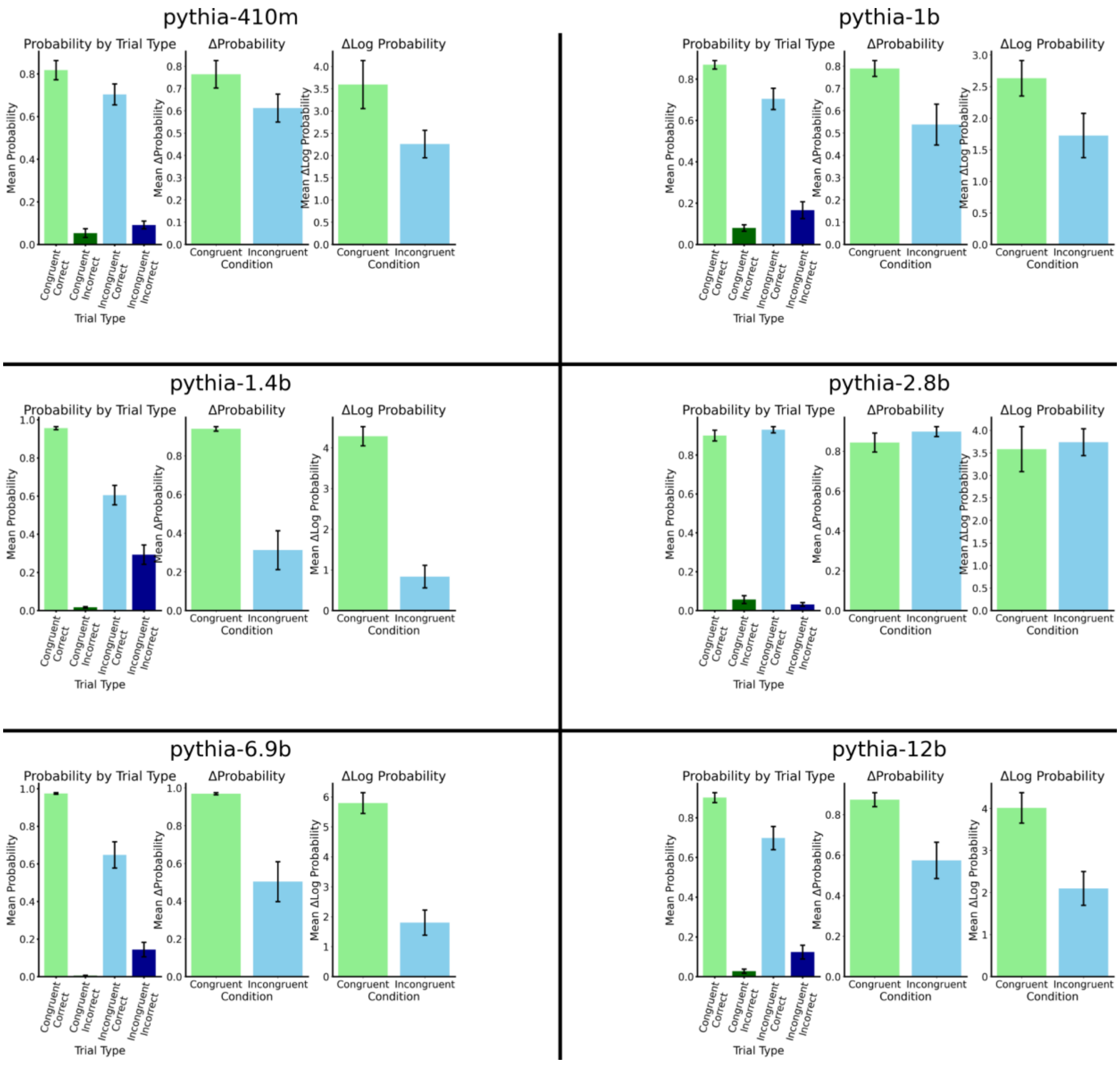

**Figure S6**: *Congruency Effects by Trial Type and Task Condition*.

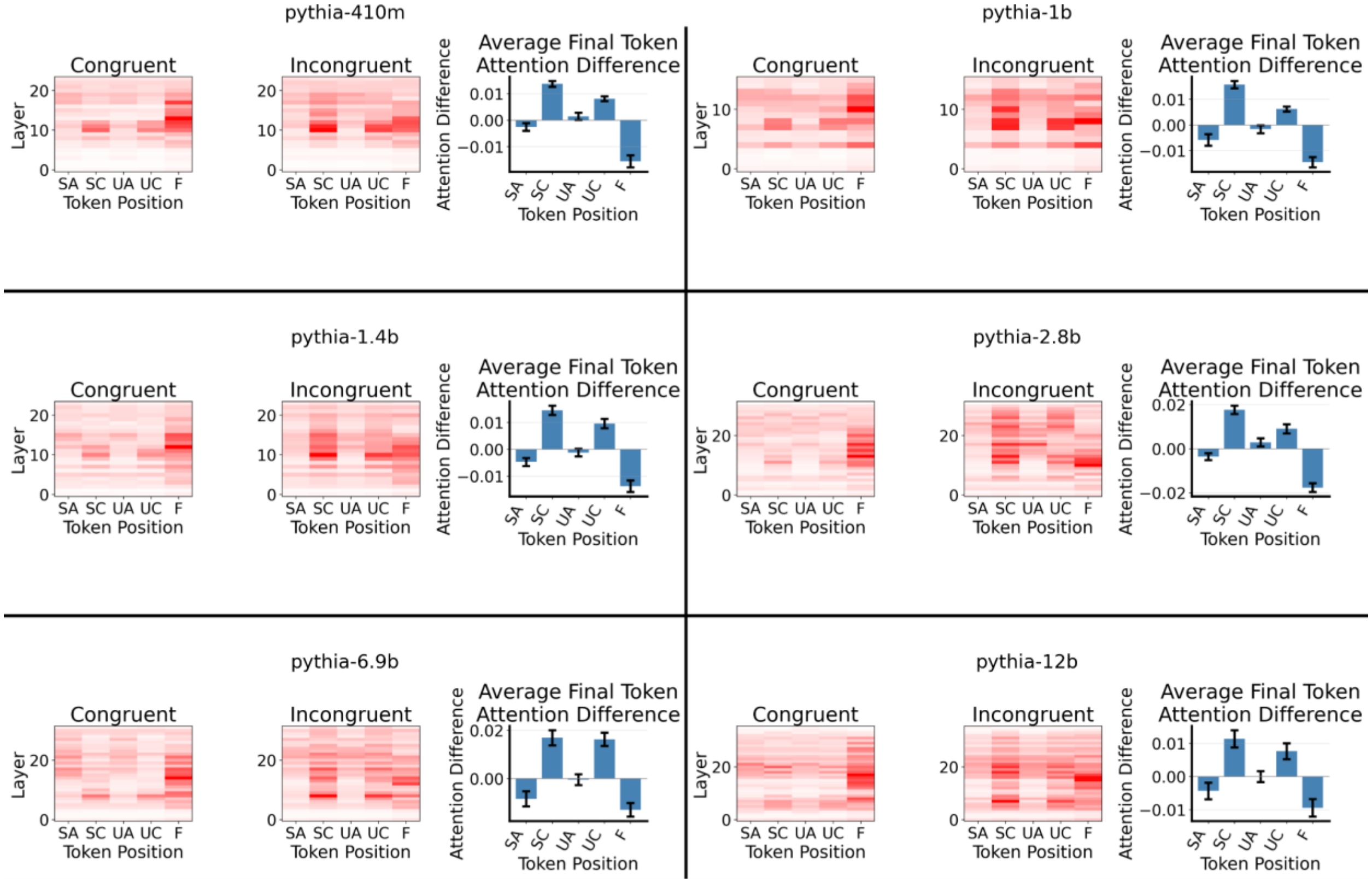

**Figure S7**: *Attention Patterns by Task Condition*.

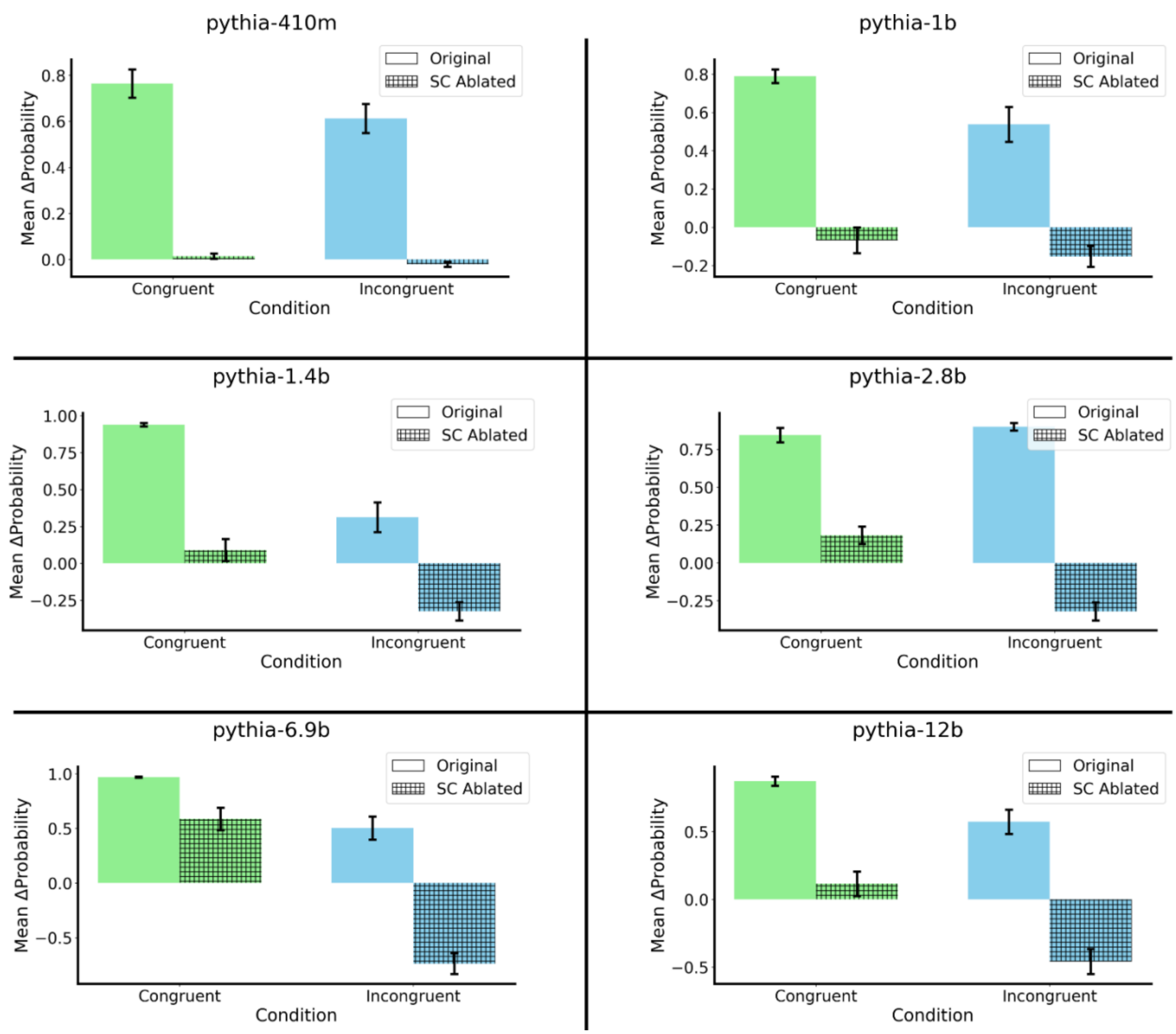

**Figure S8**: *Effects of Attention Ablation by Task Condition*.

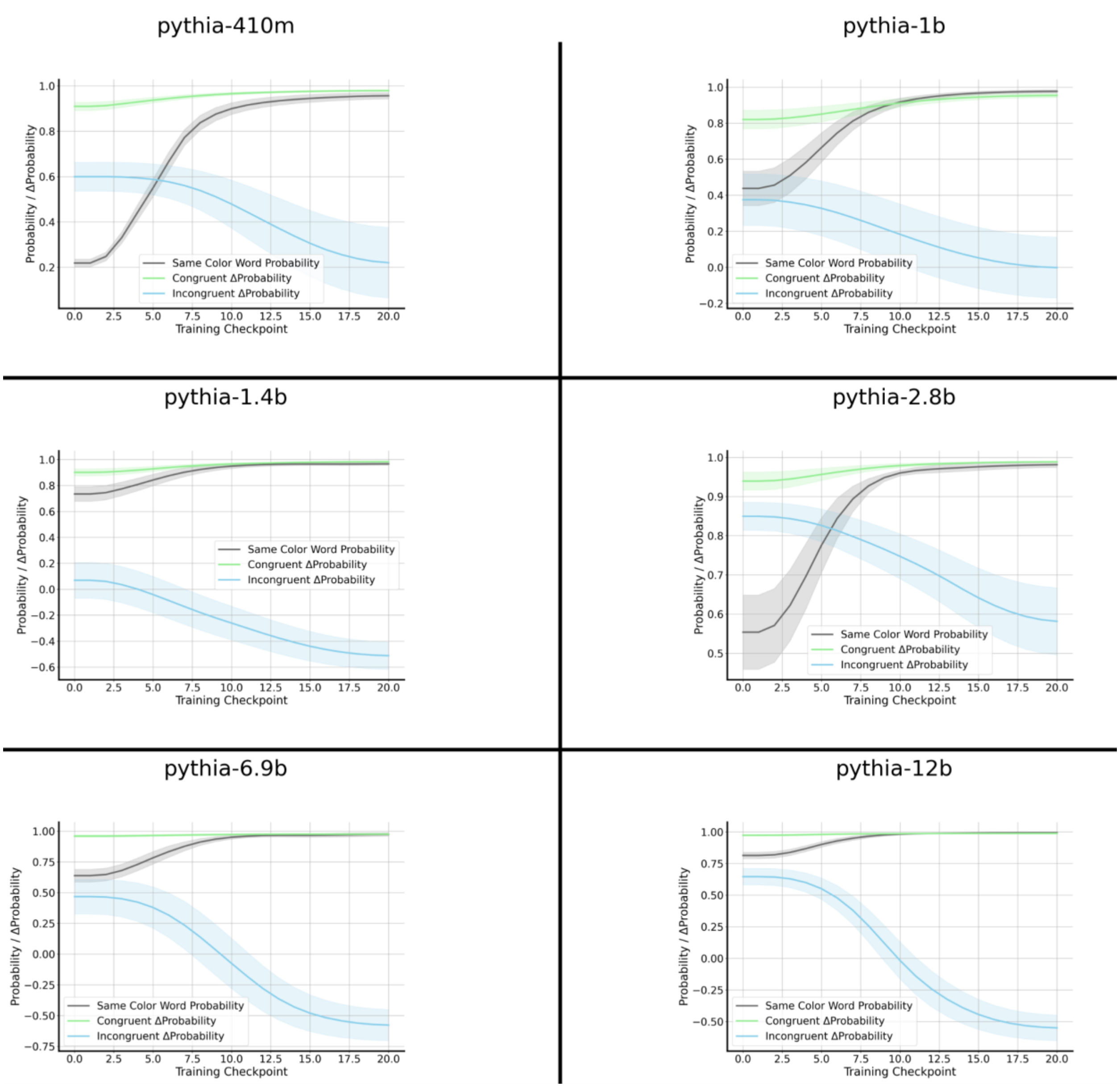

**Figure S9**: *Effects of fine tuning to amplify the default "same color word" tendency in Pythia models.*

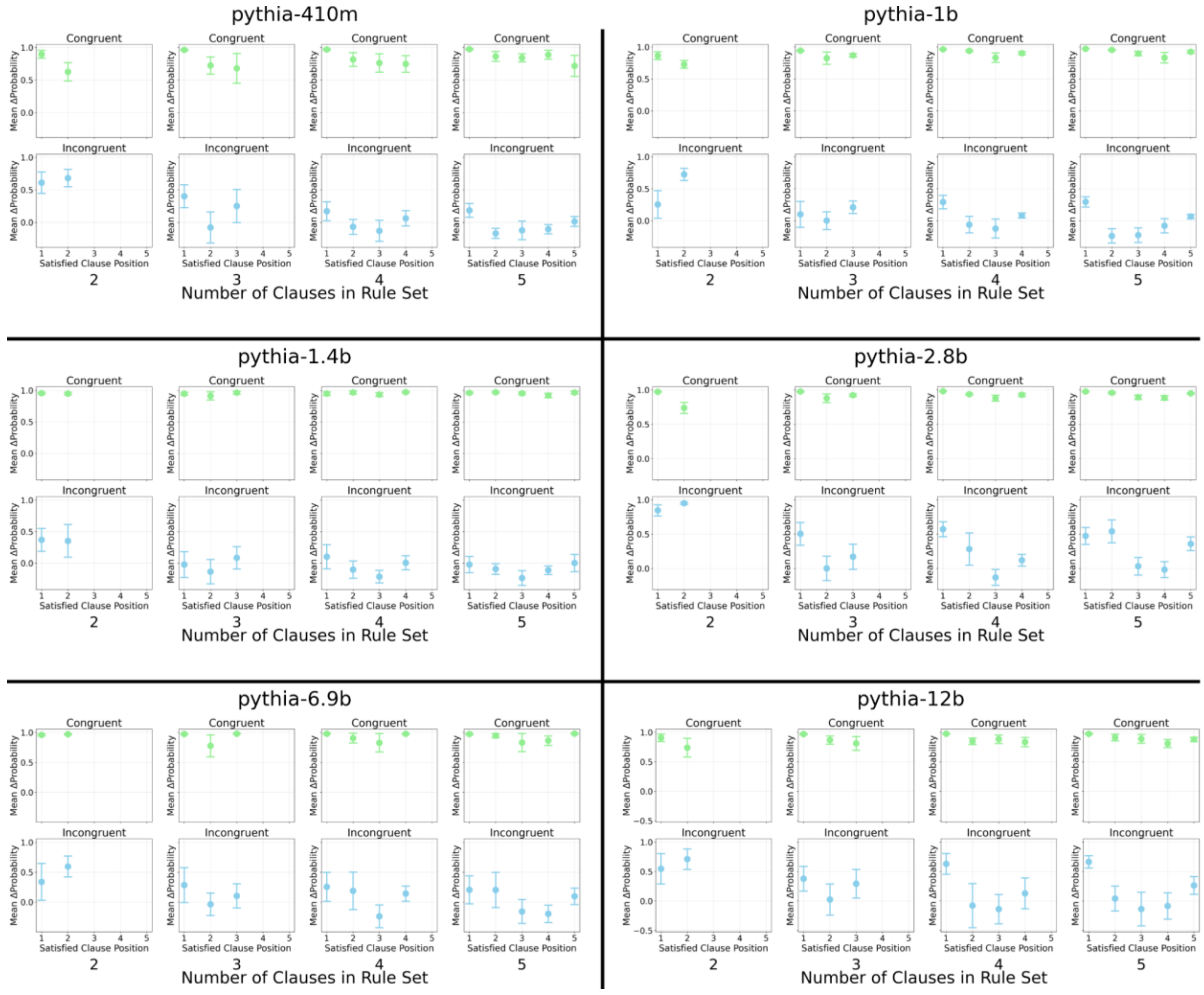

**Figure S10**: *Effects of varying the number of clauses in the rule set by task condition.*